\documentclass[12pt]{iopart}
\usepackage{iopams}
\newtheorem{defi}{Definition}
\newtheorem{lem}[defi]{Lemma}
\newtheorem{thm}[defi]{Theorem}
\newtheorem{cor}[defi]{Corollary}

\newtheorem{remark}{Remark}
\makeatletter
\newenvironment{pf}
{\noindent{\bf Proof}\quad}{\leavevmode\hfill$\square$\par\@endpetrue}
\makeatother
\def\real{\mathbb{R}}
\def\complex{\mathbb{C}}

\def\num{\mathbb{N}}
\def\su{\mathfrak{su}}
\def\sl{\mathfrak{sl}}
\def\SU{\mathop{\rm SU}}
\def\SL{\mathop{\rm SL}}
\def\SO{\mathop{\rm SO}}
\def\fdom{{\cal D}_f}
\def\odom{{\cal D}_o}

\def\Id{I}
\def\im{\mathop{\rm Im}}

\begin{document}
\jl{1}
\title[Subnormal operators regarded as generalized 
observables]{Subnormal operators \\ regarded as generalized observables \\ 
and \\ compound-system-type normal extension 
\\ related to $\su(1,1)$}
\author{Masahito Hayashi\dag and
Fuminori Sakaguchi\ddag}

\address{\dag\ 
Department of Mathematics, Graduate School of Science, Kyoto University,
Kyoto, 606-8502, Japan. e-mail  masahito@kusm.kyoto-u.ac.jp}
\address{\ddag\
Department of Electrical and Electronics Engineering, Fukui University,
3-9-1, Bunkyo, Fukui, 910-8507, Japan. e-mail saka@dignet.fuee.fukui-u.ac.jp}

\begin{abstract}
In this paper, subnormal operators, not necessarily bounded,
are discussed as generalized observables.
In order to describe not only the information about 
the probability distribution of the output data of their measurement 
but also a framework of their implementations, 
we introduce a new concept {\it compound-system-type normal extension}, 
and we derive the compound-system-type normal extension
of a subnormal operator, which is defined from an irreducible unitary 
representation of the algebra $\su(1,1)$.  The squeezed states 
are characterized as the eigenvectors of an operator 
from this viewpoint, and 
the squeezed states in multi-particle systems are shown to be 
the eigenvectors of the adjoints of these subnormal operators 
under a representation. 
The affine coherent states are discussed in the same context, as well.
\end{abstract}
\submitted
\maketitle

\section{Introduction}\label{intr}
 In quantum mechanics, observables are described by 
self-adjoint operators and the probability distributions 
of the output data of their measurement 
are determined by the spectral measures 
of those self-adjoint operators and the density operators of states.

When a linear operator has its spectral measure,
it is a {\it normal operator} where its self-adjoint part 
and its skew-adjoint part commute with each other 
(Lemma \ref{H1.1}).
In a broader sense, therefore, it can be regarded as a complexified observable.
(NB: From this viewpoint, 
in the following, we will use the expression ``measurement of
a normal operator'' in this wider sense, even if the normal 
operator is not always self-adjoint.) 
However, the measurements in quantum systems, which 
are not necessarily the measurements of any observables, 
are described by the {\it positive operator-valued measures} (POVM), 
which are a generalization of spectral measures 
(Definition \ref{D11} and Lemma \ref{15.3}).
In this paper, from these viewpoints, we try to treat 
the observables generalized even for the class of 
{\it subnormal operators}\footnote{The concept of subnormality 
was introduced by Halmos \cite{Hal1,Hal2}.}, 
which is known as a wider class including the class of normal operators.
A subnormal operator is defined as the restriction
of the normal operator into a narrower domain.
As far as the authors know, 
such a idea generalizing observables was introduced by Yuen and Lax \cite{YL}.
The pair of the normal operator and the wider domain is called
its {\it normal extension} (Definition \ref{D2}).
We can define the POVM of a subnormal operator uniquely 
in a similar sense that we can define the spectral measure of 
a normal operator uniquely under some condition. (Lemma \ref{L88}).
By this correspondence, we will formulate the measurements 
of the subnormal operators which are not necessarily bounded.
In this paper, we will not only investigate the POVMs 
of the subnormal operators 
but also give some examples of frameworks of 
their implementations in a physical sense.

There are many cases where the adjoint operator of 
a subnormal operator has eigenvectors with continuous 
potency and an over-complete eigenvector system.
In these cases, the POVM constructed from the over-complete 
eigenvector system is just the POVM of the subnormal operator
(Lemma \ref{H9}). 
Thus the subnormal operator is closely related to eigenvectors 
with continuous potency and to over-complete function systems, 
and these relations are important for the discussions on the 
properties of the subnormal operator.
This fact may give us an illusion 
that the adjoint of any operator
with a point spectrum with continuous potency 
would be a subnormal operator. 
However, the subnormality is not necessarily guaranteed only by the condition 
that its adjoint has point a spectrum with continuous 
potency\footnote{Its counter examples are given in Lemma \ref{720.1} and
Lemma \ref{720.2}.}. 

For example, an implementation of the measurement 
of a subnormal operator has been already known 
for an actual system in quantum optics.
Let $Q$ and $P$ be the multiplication operator and 
the $(-i)$-times differential operator on the Hilbert space 
$L^2(\real )$. 
A POVM is constructed from the over-complete eigenvector
system of the boson annihilation operator $a_b:= \sqrt{1/2}( Q + i P)$ 
(known as the coherent states system).
Then this POVM is just the POVM of the boson creation operator $a_b^*$
which is a subnormal operator.
The measurement of this POVM 
has been implemented as is shown in the following (see section \ref{14.2}, 
in detail), and is called the {\it heterodyne measurement};
this implementation is performed by the measurement 
of a normal operator on the compound system between 
the basic system 
(i.e. the system of interest where the measurement is 
originally discussed) and an additional ancillary system 
prepared appropriately.
Note that this operation, 
of measuring a normal operator on the compound system 
by preparing an additional ancillary system, 
gives a kind of normal extension
of creation operator $a_b^*$.
But, only giving the definition of the normal extension 
is not sufficient for discussing such a physical operation.
For clarifying such a physical operation, in section \ref{14.2}, 
we will introduce a new concept
{\it compound-system-type normal extension} which describes
not only the normal extension but also a framework of a physical operation 
(given in Definition \ref{D8}).

In section \ref{s3}, under the circumstance where an irreducible 
unitary representation of the algebra $\su(1,1)$ is given, we will 
construct two types of operators which have point spectra with 
continuous potency, and will investigate what condition guarantees 
the subnormality of these operators. 
The coherent states of the algebra $\su(1,1)$ 
introduced by Perelomov \cite{Pe}, 
will be reinterpreted as the eigenvectors of these operators. 
Moreover, in section \ref{s5}, we will derive the compound-system-type
normal extensions of these operators when they are subnormal operators.

In section \ref{s42}, from the relationship between the irreducible unitary
representations of the algebra $\su(1,1)$ and those of the affine group 
($ax+b$ group), we will discuss what subnormal operators are related to
the irreducible unitary representations of the affine group. 
Moreover, we will discuss the correspondence between the eigenvectors  
of this subnormal operator (or the coherent states
of the the algebra $\su(1,1)$ ) and the coherent states of the 
affine group. Hence we will show a relationship 
between our problem and the irreducible unitary representation of 
the affine group which is closely related to the continuous wavelet transform.

Next, in section \ref{s43}, 
from the relationship between the representation of 
the algebra $\su(1,1)$ and the squeezed states, it will be confirmed 
that the squeezed states can be described as the coherent 
states of the algebra $\su(1,1)$ in our context. 
In other words, the squeezed states are characterized as 
the eigenvectors of the operators (with point spectra 
with continuous potency) which are canonically constructed from an 
irreducible unitary representations of the algebra $\su(1,1)$. 
However, the adjoints of these operators are not necessarily
subnormal operators and are not directly regarded 
as generalized observables.

We can easily confirm that the squeezed states are the eigenvectors 
of an operator with a point spectrum with continuous potency as follows;
according to Yuen \cite{Yu}, let 
$b_{\mu,\nu}:= \mu a_b + \nu a_b^*$ with 
$|\mu|^2 - |\nu|^2 =1$, 
and characterize the squeezed state by  the eigenvector 
$|\alpha;\mu,\nu\rangle$
of the operator $b_{\mu,\nu}$ associated
with the eigenvalue $\alpha \in \complex$.
In the special cases where $\alpha =0$, the vector 
$|0 ;\mu,\nu\rangle$ can be obtained by operating the action 
of the group with the generators 
$\frac{1}{2}Q^2 ,-\frac{1}{2}P^2$ and $\frac{1}{2}(PQ+QP)$ 
upon the boson vacuum vector $| 0 ; 1 ,0 \rangle$. The algebra 
with these generators satisfies the commutation relations 
of the algebra $\su(1,1)$.
By operating $Q^{-1}$ (or $(a_b^*)^{-1}$) 
upon the characteristic equation 
$b_{\mu,\nu}|0;\mu,\nu\rangle=0$ from the left,
we have the characteristic equations
\begin{eqnarray}
Q^{-1}P | 0;\mu,\nu\rangle = i \frac{\mu+\nu}{\mu-\nu}| 0;\mu,\nu\rangle
\label{5} \\
- (a_b^*)^{-1}a_b | 0;\mu,\nu\rangle 
=  \frac{\nu}{\mu}| 0;\mu,\nu\rangle. \label{19.2}
\end{eqnarray}
In section \ref{s43}, we will derive these two equations 
again and reinterpret them from the viewpoint of the representation theory.
In this framework, the operators $Q^{-1}P$ and 
$(a_b^*)^{-1}a_b$ have point spectra with continuous potency
and they are constructed from an irreducible unitary 
representation of the algebra $\su(1,1)$ naturally.
While the adjoints of these operators are not subnormal 
operators in the case of one-particle system, 
the adjoints of these operators
are subnormal operators 
in the cases of two-particle system and multi-particle systems.
Hence we can characterize a type of physically interpretable 
states by tensor-product,
as the eigenvectors of the adjoints of subnormal operators 
in the cases of two-particle and multi-particle systems.
\par From a more general viewpoint, our investigation in this paper 
is regarded as a problem of the joint measurement between 
the self-adjoint part and the skew-adjoint part of a subnormal operator 
which do not always commute with each other. However, we should be careful 
about the difference between self-adjoint operators 
and symmetric operators in these discussions, because 
there are many delicate problems when unbounded operators 
are treated (section \ref{s51}).

In this paper,
the complex conjugate and the adjoint operator are denoted by $~^*$.
And the closure is denoted by the overline.

\section{Subnormal operator and POVM}\label{s21}
In this section, we will summarize several well-known lemmas and 
will modify them for the discussion in the following sections.
Some of the well-known lemmas will be extended for unbounded operators, 
and the proofs of the extended version will be given, as well.
In this paper, only densely defined linear operator will be discussed. 
In the following, $\odom(X)$ denotes the domain of a linear operator $X$.
A densely defined operator $X$ is called {\it closed}
if the domain $\odom(X)$ is complete with respect to the 
{\it graph norm} $\| \phi \|_{\odom(X)} 
:= \sqrt{ \|\phi\|^2 + \| X \phi \|^2 }$.
In operator theory, for two densely defined operator $X,Y$, the product 
$XY$ is defined as $\phi \mapsto X(Y(\phi))$ for any vector $\phi$
belonging to the domain $\odom(X Y): =
\{ \phi \in \odom(X) | X \phi \in  \odom(Y) \}$.
The notation $X \subset Y$ means that $\odom(X) \subset \odom (Y)$ and
$X \phi = Y \phi , \phi \in \odom(X)$.
The notation $X=Y$, also means that $X \subset Y$ and $Y \subset X$.
We will begin with reviewing the definition of normal operator 
and that of subnormal operator in unbounded case.
\begin{defi}\label{D1}\rm
A closed operator $T$ on ${\cal H}$ is called 
{\it normal} if it satisfies the condition
$T^* T = T T^*$.
\end{defi}
Remark that the operator $X^*X$ is defined on its domain 
$\odom(X^*X):=  \{ \phi \in \odom(X) | X \phi \in \odom(X^*) \}$
and it is self-adjoint and non-negative.
\begin{defi}\label{D2}\rm
A closed operator $S$ is called {\it subnormal} if there exists a Hilbert space
${\cal K}$ including ${\cal H}$ and a normal operator $T$ on ${\cal K}$
such that
$S = T P_{\cal H}$,
where $P_{\cal H}$ denotes the projection from ${\cal K}$ to ${\cal H}$
and we write the operator $S P_{{\cal H}}$ on the bigger space ${\cal K}$ 
by $S$.
In the following, we call the pair $({\cal K},T)$ a {\it normal extension} of
the subnormal operator $S$.
\end{defi}
\begin{remark}\rm
Many papers,
for example, 
Stochel and Szafraniec \cite{SSz1,SSz},
Szafraniec \cite{Sz},
\^{O}ta \cite{Ota} 
and Lahti, Pellonp\"{a}\"{a} and Ylinen \cite{LPY},
adopt another definition of
the subnormality, which substitutes $S \subset T P_{\cal H}$
for $S = T P_{\cal H}$.
According to \^{O}ta \cite{Ota},
there exists an example which is not subnormal in our definition,
but subnormal in their definition.
\end{remark}
For a spectral measure (i.e. a resolution of identity by projections)
$E$ over $\complex$,
$\int_{\complex} z E (\,d z) $ denotes the operator
$\phi \mapsto \lim_{n \to \infty} 
\left(\int_{|z | \,< n} z E (\,d z)\phi \right)
$ with the domain $\left\{ \phi \in {\cal H} \left|
\int_{\complex} |z|^2 \langle \phi, E(\,d z) \phi \rangle \,< \infty 
\right.\right\}$.
Concerning normal operators, the following lemma is well-known.
See Theorem 13.33 in Rudin \cite{Ru}.
\begin{lem}\label{H1.1}
For a normal operator $T$, there uniquely exists a spectral measure $E_T$
over $\complex$ such that 
$ T = \int_{\complex} z E_T(\,d z) $.
\end{lem}
Lemma \ref{H1.1} tells that a normal operator corresponds
to a spectral measure by one to one.
Next, we will discuss measurements in a quantum system
in order to investigate what is corresponding to Lemma \ref{H1.1}
in the case of subnormal operators.

Let ${\cal H}$ be a Hilbert space representing a physical
system of interest.
Then, the state is denoted by a non-negative operator
$\rho$ on ${\cal H}$ whose trace is $1$.
It is called a {\it density operator} on ${\cal H}$, and
the set of density operators on ${\cal H}$ is denoted by ${\cal S}({\cal H})$.
Let $P_{\rho}$ be the probability distribution given by a density 
$\rho$ and a measurement.
Then, the probabilistic property of the measurement is 
described by the map $P: \rho \mapsto P_{\rho}$.
We can naturally assume that
the map $P$ satisfies the following condition from the formulation of 
quantum mechanics:
\begin{equation}
\lambda P_{\rho_1} + (1- \lambda) P_{\rho_2} = 
P_{\lambda \rho_1+ (1-\lambda )\rho_2}
, \quad 0 \,< \forall \lambda \,< 1 ,
\forall \rho_1, \rho_2 \in {\cal S}({\cal H}).
\label{15.1} 
\end{equation}
\begin{lem}\label{15.3}
For a map $P$ satisfying (\ref{15.1}),
there uniquely exists a {\it positive operator valued measure (POVM)} 
$M$ defined in the following
which satisfies the condition
\begin{equation*}
P_{\rho}(B)= \tr M(B) \rho, \quad
\forall B \in {\cal F}(\Omega) , \forall \rho \in {\cal S}({\cal H}) .
\end{equation*}
\end{lem}
This lemma was proved by Ozawa \cite{Ozawa} in a more general framework.
For an easy proof of a finite-dimensional case, see section 6 in chapter I of 
Holevo \cite{Hol}.
This lemma guarantees that we have only to discuss POVMs
in order to describe probabilistic properties.
\begin{defi}\label{D11}\rm
Let $M$ be a map from a $\sigma$-field ${\cal F}(\Omega)$ over $\Omega$
to the set ${\cal B}_{sa}^+({\cal H})$ of bounded, self-adjoint and
non-negative operators on ${\cal H}$.
The map $M$ is called a {\it positive operator valued measure (POVM)}
on ${\cal H}$ over $\Omega$ if
it satisfies the following:
\begin{itemize}
\item 
$\displaystyle M( \emptyset ) = 0, \quad M (\Omega) = I \quad (I \hbox{: indentity op.})$ 
\item
$\displaystyle
\sum_{i} M( B_i ) = M( \cup_i B_i )
\hbox{ for } B_i \cap B_j = \emptyset, \quad (i \neq j). $
\end{itemize}
\end{defi}
A POVM $M$ is a spectral measure 
if and only if $M(B)$ is a projection for any $B$.
The following Lemma \ref{15.4} is called Na\v{i}mark's extension theorem.
For a proof, see Na\v{i}mark \cite{Naimark}, 
section 5 in chapter II in Holevo \cite{Hol} or
Theorem 6.2.18 in Hiai and Yanagi \cite{HY}.
It implies that the set of spectral measures is an important class in POVMs.
\begin{lem}\label{15.4}
Let $M$ be a POVM over a $\sigma$-field ${\cal F}(\Omega)$ on a 
Hilbert space ${\cal H}$.
There exist a Hilbert space ${\cal K}$ including ${\cal H}$ and
a spectral measure $E$ on the Hilbert space ${\cal K}$ such that
\begin{equation*}
M(B) =  P_{\cal H}  E(B) P_{\cal H}, \quad \forall B \in {\cal F}(\Omega),
\end{equation*}
where $P_{\cal H}$ denotes the projection from ${\cal K}$ to ${\cal H}$.
We call such a pair $({\cal K},E)$ a {\it Na\v{i}mark extension} of
the POVM $M$.
\end{lem}
In the following, we will treat only POVMs over the complex numbers $\complex$
whose $\sigma$-field is a family of Borel sets.
\begin{defi}\rm
A closed subspace ${\cal H}'$ of ${\cal H}$ is said to {\it reduce}
a spectral measure $E$ on ${\cal H}$,
if the projection $P_{{\cal H}'}$ to ${\cal H}'$ commutes with $E(B)$ for any 
Borel set $B$.
A Na\v{i}mark extension $({\cal K},E)$ of a POVM $M$ on ${\cal H}$
is called {\it minimal} if
${\cal K}$ has no non-trivial subspace which includes ${\cal H}$ and
reduces the spectral measure $E$.
\end{defi}
The following lemma guarantees the uniqueness of 
the minimal Na\v{i}mark extension.
It is proved as a corollary of PRINCIPAL THEOREM in section 6 of
Appendix in Riesz and Sz.-Nagy \cite{RieszN}.
\begin{lem}\label{L810}
Let $({\cal K}_1,E_1)$ and $({\cal K}_2,E_2)$ be Na\v{i}mark extensions
of a POVM $M$ on ${\cal H}$.
There exists a unitary map $V$ from ${\cal K}_1$ to ${\cal K}_2$
such that $U \phi = \phi$ for any $\phi \in {\cal H}$
and $V E_1(B) V^* = E_2(B)$ for any Borel $B$.
\end{lem}
We will give the following definition with respect to 
the inequalities among linear operators not necessarily
bounded.
\begin{defi}\label{D3}\rm
For non-negative and self-adjoint operators $X,Y$ on ${\cal H}$,
we denote $X \ge Y$ if they satisfy 
\begin{equation*}
\langle \phi , X \phi \rangle \ge \langle \phi , Y \phi \rangle 
, \quad \forall \phi \in 
\fdom(q(X)) \subset \fdom(q(Y)).
\end{equation*}
where $q(X)$ denotes the closed non-negative
quadratic form defined by a non-negative self-adjoint operator
$X$ and $\fdom(q)$ denotes the domain of a closed non-negative
quadratic form $q$.
\end{defi}
We introduce the operators 
${\rm E}(M)$ and ${\rm V}(M)$ on ${\cal H}$ which are represent formally  
$ \int_{\complex} z  M(\,d z) $ and 
$ \int_{\complex} |z|^2  M(\,d z) $, respectively. 
Later, by using Lemma \ref{L91}, 
we will give more rigorous definition of ${\rm E}(M)$ and ${\rm V}(M)$. 
Then, for $\phi \in \fdom({\rm q}(M)), \| \phi \| =1$ and a POVM $M$,
the expectation of the measurement of the state by the POVM $M$
is $\langle \phi | {\rm E}(M) | \phi \rangle $
and the variance of it is $\langle \phi | {\rm V}(M) | \phi \rangle - 
|\langle \phi | {\rm E}(M) | \phi \rangle |^2$.
It is sufficient to evaluate the operator ${\rm V}(M)$, in order to evaluate 
the variance.
But, when they are unbounded,
we should be more careful with respect to their domains.
We define the closed non-negative
quadratic form ${\rm q}(M)$ with the domain $\fdom({\rm q}(M))$ by 
\begin{eqnarray*}
{\rm q}(M)(\phi,\phi)  := 
\int_{\complex} |z|^2 \langle \phi , M(\,d z) \phi \rangle, \quad \phi \in 
\fdom({\rm q}(M)) . \\
\fdom({\rm q}(M)) :=  \left\{ \phi \in {\cal H} \left|
\int_{\complex} |z|^2 \langle \phi , M(\,d z) \phi \rangle \,< \infty 
\right. \right\} .
\end{eqnarray*}
We assume the condition that 
the set $\fdom({\rm q}(M))$ is a dense subset of ${\cal H}$.
Let ${\rm V}(M)$ be the self-adjoint operator defined by
the closed non-negative quadratic form ${\rm q}(M)$.
Next, we will define the operator $\tilde{\rm E} (M)$.
Define 
${\rm E}_R(M):= \int_{|z| \,< R } z M(\,d z)$ .
Then, the sequence $\{{\rm E}_{n} (M) \phi \}$ is a Cauchy sequence
for any $\phi \in \fdom({\rm q}(M))$,
because 
we have $\|{\rm E}_n (M)\phi  - {\rm E}_m (M)\phi\|^2
= \int_{ n \le |z| \,< m }
|z|^2 \langle \phi , M(\,d z) \phi \rangle $ for $n \,< m$.
Therefore, we can define the vector 
$\tilde{\rm E} (M) \phi := \lim_{n \to \infty} {\rm E}_n (M) \phi $.
Thus, we can define the operator $\tilde{\rm E} (M)$
on the domain $\fdom({\rm q}(M))$.
\begin{lem}\label{L91}
The operator $\tilde{\rm E} (M)$ has a closed extension.
\end{lem}From this lemma, we can define the closed operator ${\rm E}(M) $ by
the closure of the operator $\tilde{\rm E} (M)$.

\begin{pf}
Let $(E,{\cal K})$ and $P_{\cal H}$ be a Na\v{i}mark extension of $M$
and the projection from ${\cal K}$ to ${\cal H}$.
The operator $T:= \int z E(\,d z)$ is normal. From the definition of $T$,
we have $\odom(T) = \{ \phi \in {\cal K} | \int
|z|^2 \langle \phi , E(\,d z) \phi \rangle \,< \infty \}$.
Then the domain $\fdom({\rm q}(M))$ equals $\odom(T)\cap {\cal H}$.
Let $T= U |T|$ be a polar decomposition of $T$.
Since the operator $T$ is normal, we have $U |T| = |T| U$.
This equation implies that the domain of $|T|$ is invariant under 
the action of $U$.

In general, for a closed operator $X$ on ${\cal K}$ and 
closed subset ${\cal H}$ of ${\cal K}$,
the operator $X P_{\cal H}$ with the domain $\odom(X) \cap {\cal H}$
is closed if $\odom(X) \cap {\cal H}$ is dense in ${\cal H}$.
We can define the closed operator $T^* P_{\cal H}$
on its domain $\odom(T^* P_{\cal H}):= \odom(T^* ) \cap {\cal H}= \odom(T)\cap {\cal H}
= \fdom({\rm q}(M))$.
Then, we have the relation $\odom((T^* P_{\cal H})^* ) \supset \odom(T ) $.
Define the closed operator $(T^* P_{\cal H})^*P_{\cal H}$ on its domain 
$\odom((T^* P_{\cal H})^* P_{\cal H} ):= \odom((T^* P_{\cal H})^*) \cap {\cal H}
\supset \odom(T ) \cap {\cal H} = \fdom({\rm q}(M))$.
Then, we obtain $(T^* P_{\cal H})^* P_{\cal H} \supset \tilde{\rm E}(M)$.
It follows that the operator $\tilde{\rm E}(M)$ has 
a closed extension.
\end{pf}

\begin{lem}\label{H4}
Let $X$ and $M$ be an operator on a Hilbert space ${\cal H}$
and a POVM on the Hilbert space ${\cal H}$, respectively.
If $X \supset {\rm E}(M)$, then we have 
${\rm V}(M) \ge X^* X$.
\end{lem}
\begin{pf}
For a vector $\phi \in \fdom({\rm q}(M))$, we have
\begin{eqnarray*}
{\rm q}(M)(\phi,\phi) - \langle \phi | X^* X | \phi \rangle 
= \int_{\complex} \langle  \phi  | (z^*- X^*) M( \,d z )
(z-X) | \phi \rangle \ge 0.
\end{eqnarray*}
Since the relation $\fdom({\rm q}(M)) \subset \odom( {\rm E}(M)) \subset \odom(X)$ holds,
we obtain Lemma \ref{H4}.
\end{pf}
The bounded version of this lemma is proved by Helstrom \cite{Hel}
from the viewpoint of quantum estimation theory.
Its bounded version, also, follows from Kadison's inequality \cite{Kadison}.
\begin{lem}\label{H6}
Let $S$ be an operator defined on the dense subset $\odom(S)$ of ${\cal H}$.
The operator $S$ is subnormal if and only if there exists a POVM
$M$ satisfying the conditions 
\begin{eqnarray}
S = {\rm E}(M) \label{99.2} \\
S^* S = {\rm V}(M) . \label{99.3}
\end{eqnarray}
\end{lem}
\begin{pf}
Let $({\cal K},T)$ and $P_{\cal H}$ be a normal extension of the operator $S$
and the projection from ${\cal K}$ to ${\cal H}$, respectively.
By defining a POVM $M$ by $M(B):= P_{\cal H} E_T(B) P_{\cal H}$,
the equation (\ref{99.2}) is trivial.
Since the equation ${\rm V}(M)= (T P_{\cal H} )^* (T P_{\cal H}) = S^*S$
holds, we have the equation (\ref{99.3}).
Assume the equations (\ref{99.2}) and (\ref{99.3}). From 
Na\v{i}mark's extension theorem (Lemma \ref{15.4})
there exists a Na\v{i}mark extension $({\cal K},E)$ of the POVM $M$.
Define a normal operator $T:= \int_{\complex} z E(\,d z)$.
Then we have ${\rm V}(M)= (T P_{\cal H} )^* (T P_{\cal H}) , {\rm E}(M)=
P_{\cal H} (T P_{\cal H})$. From the equations (\ref{99.2}), 
(\ref{99.3}) and Lemma \ref{H5},
we can prove that $S$ is subnormal.
\end{pf}
The bounded version of this lemma is proved by Bram \cite{Bra}.
\begin{defi}\label{14.1}\rm
A POVM $M$ is called a {\it POVM} of a subnormal operator $S$ if
$M$ satisfies the preceding conditions (\ref{99.2}) and (\ref{99.3}).
\end{defi}
We will prove Lemma \ref{H5} applied in the proof of Lemma \ref{H6}.
\begin{lem}\label{H5}
Let $S$, ${\cal K}$ and $P_{\cal H}$
be an operator on a Hilbert space ${\cal H}$, a Hilbert space including 
the Hilbert space ${\cal H}$ and the projection from ${\cal K}$ to ${\cal H}$,
respectively.
For an operator $T$ on ${\cal K}$,
the following are equivalent:
\begin{description}
\item[(A)] $\displaystyle S = TP_{\cal H}$.
\item[(B)] $\displaystyle S^* S = (T P_{\cal H})^*(T P_{\cal H}),
\quad  S= P_{\cal H} (T P_{\cal H}) $.
\end{description}
\end{lem}
\begin{pf}
It is easy to derive the condition ${\bf (B)}$ from the condition ${\bf (A)}$.
Assume the condition ${\bf (B)}$.
We have 
$(T P_{\cal H})^* (T P_{\cal H} )
= (P_{\cal H}  (TP_{\cal H}))^* (P_{\cal H} (TP_{\cal H}) )
+ \left( (I -P_{\cal H}) T P_{\cal H} \right)^* 
(( I -P_{\cal H}) (T P_{\cal H}))$
and $(P_{\cal H} (TP_{\cal H}))^* (P_{\cal H} (TP_{\cal H}))=
S^* S = (T P_{\cal H})^*( T P_{\cal H})$.
Therefore, we obtain $( I -P_{\cal H})  (TP_{\cal H}) =0$.
Thus, we get the condition ${\bf (A)}$.
\end{pf}
\begin{defi}\rm
A closed subspace ${\cal H}'$ of ${\cal H}$ is said to {\it reduce}
a normal operator $T$ on ${\cal H}$, if 
the closed subspace ${\cal H}'$ of ${\cal H}$ reduce 
its spectral measure $E_T$.
This condition is equivalent to the condition that 
the projection $P_{{\cal H}'}$ commutes with the operators
$U$, $U^*$ and $e^{i t | T |^2 }$ for any real number $t$, where
$T = U |T|$ is the polar decomposition of $T$ with unitary $U$.
A normal extension $(T,{\cal K})$ of a subnormal operator $S$ on ${\cal H}$
is called {\it minimal} if
${\cal K}$ has no non-trivial subspace which includes ${\cal H}$ and
reduces the normal operator $T$.
\end{defi}
The POVM $M(B) := P_{\cal H} E_T(B) P_{\cal H}$ can be defined for 
a normal extension $(T,{\cal K})$ of a subnormal operator $S$,
and it is a POVM of $S$.
Conversely, from Lemma \ref{L810},
if the normal extension $(T,{\cal K})$ is minimal,
the spectral measure $E_T$ is unitarily equivalent with the minimal 
Na\v{i}mark extension of $M$.
Therefore, there exists a one-to-one correspondence between 
minimal normal extensions of a subnormal operator $S$
and its POVMs.
\begin{lem}\label{L13}
A normal extension $(T,{\cal K})$ of a subnormal operator $S$ on ${\cal H}$
is minimal if and only if
${\cal K} = \overline{\cal L}$ where the subspaces ${\cal L}$ and 
${\cal C}$ of ${\cal K}$
is defined as
\begin{eqnarray*}
{\cal L} : =
\left\{
\left. \sum_{k=1}^n (U^*)^{k} \psi_k \right| 
\psi_k \in \overline{\cal C}, n \in \num
\right\} \\
{\cal C} : =
\left\{\left. \sum_{k=1}^n 
e^{i t_k |T|^2 } \psi_k \right| \psi_k \in {\cal H} , t_k \in \real,
n \in \num \right\}. 
\end{eqnarray*}
where
$T = U |T|$ is the polar decomposition of $T$ with unitary $U$.
\end{lem}
\begin{pf}
Assume that a closed subspace ${\cal K}'$ of ${\cal K}$ including ${\cal H}$
reduces the normal operator $T$.
Then, for any $h \in {\cal H}$, any integer $m$ and any real number $t$,
we have $e^{i t|T|^2} h \in {\cal K}'$.
Since the closed subspace ${\cal K}'$ includes ${\cal C}$,
the closed subspace ${\cal K}'$ includes $\overline{\cal C}$.
Similarly, we can show that the closed subspace ${\cal K}'$ includes 
$\overline{\cal L}$ from this fact.

Next, we will prove that
the closed subspace $\overline{\cal C}$ is invariant for $U$.
It is sufficient to show that
$U \phi \in \overline{\cal C}$ for any $\phi \in {\cal H}$. From 
the definition of ${\cal C}$,
the closure $\overline{\cal C}$ reduces the operator $|T|^2$.
Also, it reduces the operators $|T|$ and $|T|^{-1}$.
Since $\odom(|T|^{-1}) \subset \im T$,
$U \phi = |T|^{-1} S \phi \in \overline{\cal C}$ 
holds for any $\phi \in \odom(S)$.
We have $U {\cal H} \subset \overline{\cal C}$ 
because $U$ is bounded and $\odom (S)$ is dense in ${\cal H}$.
Thus, $U e^{i t |T|^2} \phi =e^{i t |T|^2} U \phi \in 
\underline{\cal C}$ for any $\phi \in {\cal H}$.
It follows that ${\cal C}$ is invariant for $U$.

Therefore, 
we have the relations
$U \overline{\cal L} \subset \overline{\cal L}$, 
$U^* \overline{\cal L} \subset \overline{\cal L}$
and $ e^{i t |T|^2}\overline{\cal L} \subset \overline{\cal L}$
for any real number $t$.
These imply that $[ P_{\overline{\cal L}} , U] =0$,
$[ P_{\overline{\cal L}} , U^*] =0$ and
$[ P_{\overline{\cal L}} , e^{i t |T|^2} ] =0$.
It follows that the closed subspace $\overline{\cal L}$
reduces the normal operator $T$.
The lemma is now immediate.
\end{pf}
\begin{lem}\label{L14}
Let $(T,{\cal K})$ be a minimal normal extension of 
a subnormal operator $S$ on ${\cal H}$.
A Hilbert space ${\cal K}'$ including ${\cal H}$ 
and a normal operator $T'$ satisfy the condition 
$S \subset T' P_{\cal H}$. 
The following three conditions are equivalent.
\begin{description}
\item[(A)] $\langle \phi_1, e^{i t|T|^2} \phi_2 \rangle 
= \langle \phi_1, e^{i t|T'|^2} \phi_2 \rangle $ holds for any
$\phi_1, \phi_2 \in {\cal H}$.
\item[(B)] $\langle \phi_1, e^{i t|T|} \phi_2 \rangle 
= \langle \phi_1, e^{i t|T'|} \phi_2 \rangle $ holds for any
$\phi_1, \phi_2 \in {\cal H}$.
\item[(C)] 
There exists an isometric map $V$ from ${\cal K}$ to ${\cal K}'$ such that
$V \phi = \phi $ for any $\phi \in {\cal H}$ and 
$V T V^* = T'P_{\im V}$.
\end{description}
The condition ${\bf (C)}$ implies that
$T' P_{\cal H} = S$ i.e. the pair $(T',{\cal K}')$ is a normal extension 
of $S$.
\end{lem}

\begin{pf}
It is easy to show that 
the conditions ${\bf (A)}$ and ${\bf (B)}$ 
follows from the condition ${\bf (C)}$.
First, we prove that the condition ${\bf (A)}$ implies 
the condition ${\bf (B)}$.
Define the subspace ${\cal C}'$ of ${\cal K}'$ by
$\displaystyle
{\cal C}' := \left\{\left. \sum_{k=1}^n 
e^{i t_k |T'|^2 } \psi_k \right| \psi_k \in {\cal H} , t_k \in \real,
n \in \num \right\}$.

Similarly to Proof of Lemma \ref{L13},
we can prove that the closure $\overline{\cal C}$ reduces
$|T|^2$ and the closure $\overline{{\cal C}'}$ reduces
$|T'|^2$.
Then, the closures $\overline{\cal C}$ and $\overline{{\cal C}'}$ reduce
the operators $|T|$ and $|T'|$, respectively. From the condition ${\bf (A)}$,
$\langle e^{i t_1 |T|^2} \phi_1, e^{i t_2 |T|^2} \phi_2 \rangle =
\langle e^{i t_1 |T|^2} \phi_1, e^{i t_2 |T'|^2} \phi_2 \rangle $
holds for any $\phi_1, \phi_2 \in {\cal H}$ and any real numbers $t_1,t_2$.
Therefore,
we can define the unitary map $V_{\cal C}$ from 
$\overline{\cal C}$ to $\overline{\cal C}'$ by 
\begin{equation*}
V_{\cal C} \left( \sum_{k=1}^n e^{i t_k |T|^2 } \phi_k
\right) = \sum_{k=1}^n e^{i t_k |T'|^2 } \phi_k .
\end{equation*}
Thus, we have 
$V_{\cal C} |T|^2 V_{\cal C}^* = |T'|^2 $ on $\overline{{\cal C}'}$.
It implies that $V_{\cal C} |T| V_{\cal C}^* = |T'| $ 
on $\overline{{\cal C}'}$
because the closures $\overline{{\cal C}}$ and $\overline{{\cal C}'}$ 
reduce the operators 
$|T|$ and $|T'|$, 
respectively.
Since 
$V_{\cal C} \phi = \phi $ for any $\phi \in {\cal H}$
and 
$V_{\cal C} e^{it |T|} V_{\cal C}^* = e^{i t |T'|} $ for any $t \in
\real$,
the condition ${\bf (B)}$ holds.
Similarly, we can prove that the condition ${\bf (B)}$ implies 
the condition ${\bf (A)}$.

Next, we prove that the condition ${\bf (A)}$ implies 
the condition ${\bf (C)}$. From the above discussion, 
we can define the inverses $|T|^{-1}$ and $|T'|^{-1}$ on
$\im | T | \cap {\overline{\cal C}}$ and 
$\im | T' | \cap {\overline{\cal C'}}$, respectively.
Then we have $V_{\cal C} |T|^{-1}V_{\cal C}^* = |T'|^{-1}$ on 
$\im | T' | \cap {\overline{\cal C'}}$.

Let $T= U |T|$ and $ T' = U' |T'|$ be the polar decompositions of $T$
and $T'$ satisfying that $U$ and $U'$ are unitary, respectively.
The image $\im |T |$ is invariant under the unitary transformation $U$,
and the image $\im |T' |$ is 
invariant under $U'$.
Then, we have
$\im S \subset \im |T| \cap {\cal H} $ 
Similarly, we have $\im S \subset \im | T' | \cap {\cal H}$.
Thus, for any $\phi \in \im S$,
we have $V_{\cal C} |T|^{-1} \phi = |T'|^{-1} \phi$. From 
the proof of Lemma \ref{L13}, the closed subspaces
$\overline{\cal C}$ and $\overline{\cal C}$' are invariant for
$U$ and $U'$, respectively.
For any $\phi_1, \phi_2 \in \odom (S)$,
we have
\begin{eqnarray*}
\fl \langle e^{i t |T|^2} \phi_1 , U \phi_2
\rangle
&=& \langle V_{\cal C}e^{i t |T|^2} \phi_1 , V_{\cal C} U \phi_2
\rangle 
= \langle V_{\cal C} e^{i t |T|^2} \phi_1 , 
V_{\cal C}| T |^{-1} S \phi_2 \rangle \\
&=& \langle V_{\cal C} e^{i t |T|^2} V_{\cal C}^*V_{\cal C}\phi_1 , 
V_{\cal C}| T |^{-1} V_{\cal C}^*V_{\cal C}S \phi_2 \rangle 
= \langle e^{i t |T'|^2} V_{\cal C}\phi_1 , 
| T' |^{-1} V_{\cal C}S \phi_2 \rangle \\
&=& \langle e^{i t |T'|^2} \phi_1 , 
| T' |^{-1} S \phi_2 \rangle 
=  \langle e^{i t |T'|^2} \phi_1 ,  U' \phi_2\rangle .
\end{eqnarray*}
Since $e^{i t |T|^2},e^{i t |T'|^2},U$ and $U'$ are bounded,
\begin{equation*}
\langle e^{i t |T|^2} \phi_1 , U \phi_2\rangle=
\langle e^{i t |T'|^2} \phi_1 ,  U' \phi_2\rangle 
\end{equation*}
holds, for any $\phi_1, \phi_2 \in {\cal H}$.
Also, we can prove 
\begin{equation*}
\fl\left\langle \left( \sum_{k=1}^n e^{i t_k |T|^2 } \psi_k \right),
U \left( \sum_{k=1}^n e^{i t_k' |T|^2 } \psi_k' \right) \right\rangle
=
\left\langle \left( \sum_{k=1}^n e^{i t_k |T'|^2 } \psi_k \right),
U' \left( \sum_{k=1}^n e^{i t_k' |T'|^2 } \psi_k' \right)
\right\rangle,
\end{equation*}
for arbitrary $\psi_k,\psi_k' \in {\cal H}, t_k , t_k' \in \real$.
Therefore,
\begin{equation*}
\langle \phi_1 , U \phi_2 \rangle=
\langle V_{\cal C}\phi_1 , U' V_{\cal C}\phi_2 \rangle 
\end{equation*}
holds for any $\phi_1,\phi_2 \in \overline{\cal C}$.
Since 
the closed subspace
$\overline{\cal C}$ is invariant for
$U$ and the operator $U$ is bounded,
\begin{equation}
\langle \phi_1 , U^n \phi_2 \rangle=
\langle V_{\cal C}\phi_1 , {U'}^n V_{\cal C}\phi_2 \rangle \label{810.3}
\end{equation}
holds for any $\phi_1,\phi_2 \in \overline{\cal C}$ and any $n \in \num$.

We can define the isometric map $V$ from ${\cal K}= \overline{\cal L}$ to 
${\cal K'}$ by
\begin{equation*}
V\left(
\sum_{k=1}^n ({U}^*)^k \psi_k \right) =
\left(
\sum_{k=1}^n ({U'}^*)^k \psi_k \right),
\end{equation*}
where $\psi_k$ is an arbitrary element of $\overline{\cal C}$.
It can be confirmed that this definition is well-defined from
(\ref{810.3}).
Now, we can easily check the condition ${\bf (C)}$.
\end{pf}

\begin{defi}\rm
A vector $\phi \in {\cal D}^{\infty}(X):=\cap_{n=0}^{\infty}
\odom(X^n)$ 
is called an {\it analytic} vector of $X$ if 
\begin{equation}
\sum_{i=0}^{\infty}\frac{t^i}{i \!}\| X ^i \phi \| \,< \infty, 
\label{88.1}
\end{equation}
for any $t \in \real$.
The set of all analytic vectors of $S$ is written by ${\cal A}(S)$.
\end{defi}
\begin{lem}\label{L88}
Assume that 
the set ${\cal A}(S)$ is dense in ${\cal H}$
for a subnormal operator $S$ on ${\cal H}$.
Let $(T,K)$ be a normal normal extension of $S$.
If a Hilbert space ${\cal K}'$ including ${\cal H}$ 
and a normal operator $T'$ satisfy the condition 
 $T' P_{\cal H} \supset S$,
there exists an isometric map $V$ from ${\cal K}$ to ${\cal K}'$ such that
$V \phi = \phi $ for any $\phi \in {\cal H}$ and 
$V T V^* = T'P_{\im V}$.
This implies that
$T' P_{\cal H} = S$ i.e. the pair $(T',{\cal K}')$ is a normal extension 
of $S$.
Therefore, this assumption guarantees 
the uniqueness of the minimal normal extension.
\end{lem}
This lemma shows that under the assumption,
the pair $(T,{\cal H})$ is 
a normal extension of $S$ 
if a normal operator $T$ on ${\cal K}$ including ${\cal H}$ satisfies 
$T P_{\cal H} \supset S$.
For a simple proof in the bounded case, see section 2 in chapter II of Conway 
\cite{sub}.
Szafraniec \cite{Sz}
shows the uniqueness of the minimal normal extension under another
assumption that any vector $\phi \in  {\cal D}(S)$ satisfies
(\ref{88.1}) for some a real number $t \,> 0$.
Stochel and Szafraniec\cite{SSz}
discuss different sufficiently conditions for 
the uniqueness of the minimal normal extension.
\begin{pf}
It is sufficient to show that 
the condition ${\bf (A)}$ in Lemma \ref{L14} holds.
Schwarz's inequality guarantees 
\begin{equation*}
\sum_{k=0}^\infty \frac{t^k}{k !} \langle S^k \phi_1 , S^k \phi_2 \rangle
\,<\sqrt{\sum_{k=0}^\infty \frac{t^k}{k !} \| S^k \phi_1 \|^2}
\sqrt{\sum_{k=0}^\infty \frac{t^k}{k !} \| S^k \phi_2 \|^2}
\,< \infty .
\end{equation*}
for any $\phi_1 , \phi_2 \in {\cal A}(S)$ and any real number
$t$. From Fubini's Theorem,
\begin{eqnarray}
\fl \langle \phi_1 , e^{i t |T|^2 } \phi_2 \rangle 
&=& \left\langle \phi_1 , 
\sum_{k=0}^\infty \frac{(i t |T|^2)^k }{k !} \phi_2 \right\rangle
= \left\langle \phi_1 , 
\sum_{k=0}^\infty \frac{(i t)^k (T^*)^k T^k }{k !} \phi_2 \right\rangle  
\nonumber \\
&=& \sum_{k=0}^\infty  \frac{(i t)^k }{k !}
\langle S^k \phi_1 , S^k\phi_2 \rangle 
= \langle \phi_1 , e^{i t |T'|^2 } \phi_2 \rangle. \label{1.20}
\end{eqnarray}From (\ref{1.20}) and the fact that the operators 
$e^{i t |T|^2 }$ and $e^{i t |T'|^2 }$ are bounded
and ${\cal A}(S)$ is dense in ${\cal H}$,
we have the equation $
P_{\cal H} e^{i t |T|^2 } P_{\cal H}=
P_{\cal H} e^{i t |T'|^2 } P_{\cal H}$.
Therefore, the condition ${\bf (A)}$ in Lemma \ref{L88} holds.
\end{pf}From the one-to-one correspondence 
between POVMs of a subnormal operator $S$
and its minimal normal extensions,
we have the following corollary.
\begin{cor}
For any subnormal operator $S$ satisfying the assumption of Lemma 
\ref{L88},
there uniquely exists the POVM $M$ satisfying the equations (\ref{99.2}) and
(\ref{99.3}).
\end{cor}
Subnormal operators have the following properties:
\begin{lem}\label{H7}
Let $S$ be a subnormal operator on ${\cal H}$.
Then 
\begin{equation}
S^* S \ge S S^*  \label{13.2} . 
\end{equation}
\end{lem}
\begin{pf}
We have $(P_{\cal H} T) (P_{\cal H} T)^* \ge 
(P_{\cal H}(P_{\cal H} T)) 
(P_{\cal H}(P_{\cal H} T))^* = S S^*$. From the normality of $T$,
we have $S^* S = (T P_{\cal H})^*(T P_{\cal H})=
(T^* P_{\cal H})^*(T^* P_{\cal H})$.
Since $T^* P_{\cal H}\subset (P_{\cal H}T)^*$ 
(See Theorem 13.2 in Rudin\cite{Ru}),
the inequality $(T^* P_{\cal H})^*(T^* P_{\cal H}) \ge 
(P_{\cal H}T)(P_{\cal H}T)^*$ holds.
Therefore, $S^* S \ge (P_{\cal H}T)(P_{\cal H}T)^* \ge S S^* $.
\end{pf}
Operators satisfying (\ref{13.2})
are called {\it hyponormal} 
operators\footnote{This class was introduced by Halmos \cite{Hal1}.} 
and the class of these operators is important in the operator theory.
The following Lemma \ref{H9} shows
a relation between the POVM of a subnormal operator and an 
over-complete eigenvector system.
\begin{lem}\label{H9}
Let $J$ and $K$ be an operator on ${\cal H}$ and a subset of complex numbers
$\complex$, respectively.
Assume that there exists a vector $| z \rangle \in \odom(J)$ satisfying 
$J|z \rangle = z | z \rangle$ for any complex number $z \in K$,
and there exists a measure $\mu $ on $K$ satisfying 
$\int_K | z^* \rangle \langle z^*| \mu (\,d z) = I $.
Then, $J^*$ is subnormal and the POVM
$| z^* \rangle \langle z^*| \mu (\,d z)$ is 
the POVM of the subnormal operator $J^*$.
\end{lem}
\begin{pf}From the assumptions, we have
\begin{equation*}
J^*  =  \int_K | z^* \rangle \langle z^*| \mu (\,d z) J^*=
\int_K z | z^* \rangle \langle z^*| \mu (\,d z) .
\end{equation*}
Note that $\langle z^*|J^*= z \langle z^*|$.
Thus,
The POVM $M(\,d z):=| z^* \rangle \langle z^*| \mu (\,d z)$ satisfies 
the condition (\ref{99.2}).
Therefore, we obtain 
\begin{equation*}
J J^* =
\int_K  J | z^* \rangle \langle z^*| J^* \mu (\,d z)
=\int_K  |z|^2 | z^* \rangle \langle z^*| \mu (\,d z)
= {\rm V}(M).
\end{equation*}
Then the POVM $M$ satisfies the condition (\ref{99.3})
and the operator $J^*$ is subnormal, as shown from Lemma \ref{H6}.
We can confirm that the POVM $| z^* \rangle \langle z^*| \mu (\,d z)$ 
is the POVM of the subnormal operator $J^*$.
\end{pf}
In the following of this section, we treat a relation between 
a subnormal operator and its spectrum.
\begin{lem}\label{C13}
Let $S$ and $\phi$ be a subnormal operator and an eigenvector of $S$, 
respectively.
Then, a vector $\phi$ is an eigenvector of the adjoint $S^*$ operator of $S$.
\end{lem}
\begin{pf}
Let $({\cal K},T)$ and $P_{\cal H}$ be a normal extension of $S$ and
the projection from ${\cal K}$ to ${\cal H}$, respectively.
Assume that $\phi \in \odom(S)$ is an 
eigenvector of $S$ associated with an eigenvalue $c$ such that
$\| \phi \|=1$.
Since the equation $T \phi = c \phi$ holds, we have 
$T^* \phi = c^* \phi$.
Thus $S^*= P_{\cal H} T^*$.
Therefore we get $S^* \phi = c^* \phi$.
Now, we obtain the Lemma.
\end{pf}

\begin{defi}\label{D19}\rm
A subnormal operator $S$ is called {\it pure subnormal} if it 
satisfies the following condition;
if a subspace ${\cal I}$ of ${\cal H}$ satisfies 
that $S P_{{\cal I}}$ is subnormal,
then the subspace ${\cal I}$ is $\{ 0 \}$ or ${\cal H}$
\end{defi}

\begin{lem}\label{H12}
Any pure subnormal operator $S$ has no point spectrum.
\end{lem}
\begin{pf}
Let $({\cal K},T)$, $P_{\cal H}$ and $\phi$ be defined 
in Proof of Lemma \ref{C13}.
Since we have $S^* \phi = c^* \phi$,
the operator $| \phi \rangle \langle \phi|$ 
commutes with the pure subnormal operator $S$.
The fact contradicts the definition of pure subnormal operators.
\end{pf}
According to Conway \cite{sub},
it is sufficient to assume the purity and hyponormality
in Lemma \ref{H12}.

\section{Compound-system-type normal extension}\label{14.2}
Now, as an example of 
a subnormal operator and its normal extension, 
we will treat the boson creation operator $a_b^*$ 
and the {\it heterodyne measurement} in quantum optics.
The pair
$( L^2(\real ) \otimes L^2(\real ), 
a_b^* \otimes I + I \otimes a_b)$ 
is a normal extension of the subnormal operator $a_b^*$ 
under the isometric embedding 
$L^2(\real ) \to L^2(\real ) \otimes L^2(\real )$ 
defined by $\psi \mapsto \psi \otimes | 0 ; 1,0\rangle$,
where $| 0 ; 1,0\rangle$ denotes the boson vacuum vector.
Here, $a_b^* \otimes I + I \otimes a_b$ is a normal operator,
and we have $(a_b^* \otimes I + I \otimes a_b)\phi \otimes | 0 ; 1,0\rangle
= (a_b^*\phi)\otimes | 0 ; 1,0\rangle$ for any $\phi \in L^2(\real )$.
By substituting $a_b$ for $J$ in Lemma \ref{H9}    
and and by letting $| \alpha ; 1,0\rangle$ be the boson coherent state, 
we can confirm that $| \alpha ; 1,0\rangle \langle \alpha; 1,0 | 
\,d^2 \alpha$ 
is the POVM of the subnormal operator $a_b^*$.
The set of rapidly decreasing $C^\infty$ functions is dense in
$L^2(\real)$ and any rapidly decreasing $C^\infty$ function
is analytic of $a_b^*$.
Therefore, $a_b^*$'s POVM is uniquely determined.

The heterodyne measurement is implemented 
by the measurement of 
$a_b^* \otimes I + I \otimes a_b$ 
(i.e. the simultaneous measurement between 
$Q \otimes I + I \otimes Q$ and $P \otimes I - I \otimes P$ 
which commute with each other) under the circumstance where 
the state of basic system is $| \phi \rangle \langle \phi |$ and
the state of 
the ancillary system is controlled 
to be the vacuum states $| 0 ;1,0\rangle \langle 0; 1,0 |$.
In detail, see section 6 in chapter III in Holevo \cite{Hol}
or section 6 in chapter V in Helstrom \cite{Hel}.
We will generalize normal extensions of similar type 
to this, by the name of {\it compound-system-type normal extensions}, 
as follows;
\begin{defi}\label{D8}\rm
Let $S$ be a subnormal operator defined on a dense linear subspace 
$\odom(S)$ of ${\cal H}$
and let ${\cal H}'$, $T$ and $\psi$ be a Hilbert space, 
a normal operator defined 
on a dense subspace $\odom(T)$ 
of the Hilbert space ${\cal H} \otimes {\cal H}'$ and 
an element of ${\cal H}'$ whose norm is unity, respectively.
We call the triple $( {\cal H}', T ,\psi)$
a {\it compound-system-type normal extension} of the subnormal operator $S$
if it satisfies the condition
\begin{equation}
\odom(S) \otimes \psi \subset \odom(T), \quad 
\left( S \phi \right) \otimes \psi =  T \left( \phi \otimes \psi \right) 
,\quad \hbox{ for any } \phi \in \odom(S) . \label{99.1}
\end{equation}
\end{defi}
Thus the definition of the compound-system-type normal extension
describes not only the probability distribution
but also a framework of the concrete implementation process, 
while the definition of the normal extension 
given in section \ref{s21} describes only 
the probability distribution.
Therefore, a compound-system-type normal extension contains 
more informations than the corresponding POVM. 

In the following of this section,
we discuss compound-system-type normal extensions of isometric operators and 
symmetric operators, 
where we let $\{ |\uparrow \rangle, | \downarrow \rangle \}$ be a CONS of $\complex^2$.
\begin{lem}\label{H10.1}
An isometric operator $U$ defined on ${\cal H}$ is subnormal.
Define the operator $T:=
U \otimes | \uparrow \rangle \langle \uparrow | + 
U^* \otimes|  \downarrow \rangle \langle \downarrow |
+P_{{\im U}^{\perp}} \otimes | \uparrow \rangle  \langle \downarrow |$,
where ${\im U}^{\perp}$ denotes the orthogonal complementary space of $\im U$.
Then, the operator $T$ is unitary on ${\cal H} \otimes \complex^2$
and the triple $(\complex^2, T , | \uparrow \rangle )$ is 
a compound-system-type normal extension of $U$.
\end{lem}
\begin{pf}From the definition,
we have 
\begin{eqnarray*}
  T^* T 
&=&
\left(
U^* \otimes| \uparrow \rangle \langle \uparrow | 
+U \otimes|  \downarrow \rangle \langle \downarrow |
+P_{{\im U}^{\perp}} \otimes| \downarrow \rangle  \langle \uparrow | \right)\\
&& \quad \left(U \otimes| \uparrow \rangle \langle \uparrow |  
+U^* \otimes | \downarrow \rangle \langle \downarrow |
+P_{{\im U}^{\perp}} \otimes| \uparrow \rangle  \langle \downarrow |\right) \\
&=&
\Id_{\cal H} \otimes| \uparrow \rangle \langle \uparrow |  
+P_{{\im U}} \otimes | \downarrow \rangle \langle \downarrow |
+  P_{{\im U}^{\perp}} 
\otimes| \downarrow \rangle \langle \downarrow | 
= \Id_{\cal H} \otimes \Id_{\complex^2}.
\end{eqnarray*}
\begin{eqnarray*}
  T T^* 
&=&
\left(
U \otimes| \uparrow \rangle \langle \uparrow | 
+U^* \otimes|  \downarrow \rangle \langle \downarrow |
+P_{{\im U}^{\perp}} \otimes| \uparrow \rangle  \langle \downarrow | \right)\\
&& \quad \left(U^* \otimes| \uparrow \rangle \langle \uparrow |  
+U \otimes | \downarrow \rangle \langle \downarrow |
+P_{{\im U}^{\perp}} \otimes| \downarrow \rangle  \langle \uparrow |\right) \\
&=&
P_{{\im U}} \otimes| \uparrow \rangle \langle \uparrow |  
+\Id_{\cal H} \otimes | \downarrow \rangle \langle \downarrow |
+  P_{{\im U}^{\perp}} 
\otimes| \uparrow \rangle \langle \uparrow | 
= \Id_{\cal H} \otimes \Id_{\complex^2}.
\end{eqnarray*}
Then, the operator $T$ is unitary.
Moreover,
we have 
$ T ( \phi \otimes | \uparrow \rangle ) 
= ( U \phi ) \otimes | \uparrow \rangle$.
Therefore, the triple
$(\complex^2, T , | \uparrow \rangle )$ is 
a compound-system-type normal extension of $U$.
\end{pf}
A closed symmetric operator $X$ is called {\it maximal symmetric},
if there exists no symmetric operator $Y$ such that
$X \subsetneqq Y$.
\begin{lem}\label{H10}
A closed symmetric operator $X$ is subnormal on ${\cal H}$.
Define the operator $T:= X^* \otimes | - \rangle \langle + | + 
X \otimes | + \rangle \langle - |$
on the domain $\odom(T):= \odom(X^*) \otimes | + \rangle \oplus 
\odom(X) \otimes | - \rangle$ with
$| \pm \rangle := \frac{1}{\sqrt{2}} (|  \uparrow\rangle \pm | \downarrow
 \rangle)$, for the maximal symmetric operator $X$ on ${\cal H}$. 
Then, $T$ is a self-adjoint operator
and the triple $(\complex^2, T , | \uparrow \rangle )$ is 
a compound-system-type normal extension of $X$.
\end{lem}
The classification of (2nd) self-adjoint extensions of symmetric
operators is given in section 5 in Na\v{i}mark \cite{Nai1}.

\begin{pf}
We can confirm that $T$ is self-adjoint.
$\odom(T) \cap {\cal H}\otimes | \uparrow \rangle 
= \odom(X) \otimes | \uparrow \rangle $ and
$T (\phi \otimes |\uparrow \rangle )
= 
(X \phi) \otimes ( | - \rangle \langle + | + | + \rangle \langle - |)  
| \uparrow \rangle = 
(X \phi )\otimes | \uparrow \rangle
= X \phi \otimes |\uparrow \rangle $ 
holds for any $\phi \in \odom(X)$.
The lemma is immediate.
\end{pf}
For example, we apply the inequalities (\ref{13.2}) in Lemma \ref{H7}
to a symmetric operator.
If $X$ is self-adjoint, we have $X^*X= X X^*$.
But, if the operator $X$ have no self-adjoint extension,
we have $ X^*X \subsetneqq X X^*$.
This fact isn't contradictly to the inequalities (\ref{13.2}).

We have the following lemma 
from the classification by Na\v{i}mark and
the following fact; 
any maximal symmetric operator is unitarily 
equivalent with $(\Id \otimes P^+) \oplus Y$ or $(\Id \otimes P^-)
\oplus Y$, where
$Y$ is a self-adjoint operator and $P^+$ and $P^-$ are the momentum operators
on $L^2(\real^+)$ and $L^2(\real^-)$, respectively.
This fact follows from section 104 in Ahkiezer and Glazman \cite{AG}.
\begin{lem}\label{L810.2}
Any minimal normal (self-adjoint) 
extension of a closed symmetric operator $X$ is unitarily 
equivalent with each other
if and only if $X$ is maximal symmetric.
\end{lem}
\begin{remark}\rm
Lemma \ref{L810.2} gives an example of subnormal operator such that
its minimal normal extension is not unique in the sense of unitary 
equivalence.
\end{remark}
\section{Irreducible unitary representations of the algebra $\su(1,1)$ 
and their coherent states}\label{s3} 
In this section, from the minimal-weight-type 
unitary representations of the algebra $\su(1,1)$ 
(defined in this section), 
we will construct the corresponding 
subnormal operators canonically, and will investigate 
the relationship between the coherent states 
defined by Perelomov \cite{Pe} and these subnormal operators.

\begin{defi}\label{719.2}\rm
A triplet $(E_0,E_+,E_-)$ of skew-adjoint operators is called 
a {\it unitary representation of the algebra $\su(1,1)$}
if the relations
\begin{equation} 
[ E_0, E_\pm ] = \pm 2 E_\pm,  \quad [ E_+, E_- ] = E_0
\label{6}
\end{equation}
hold.
\end{defi}
For the reason for this definition, see Remark \ref{719.1}.
However, 
it is difficult to discuss the unitary representation in this
notation
because three operators $E_0,E_+,E_-$ have no eigenvector.
Thus, we define another triplet $(L_0, L_+, L_-)$ by
\begin{equation}
L_0 := i(E_- - E_+), \quad 
L_\pm := \frac{1}{2}\bigl(E_0 \pm i(E_+ + E_-)\bigr).
\label{7}
\end{equation}
Then, this triplet satisfies the commutation relations of the same type 
\begin{equation} 
[ L_0, L_\pm ] = \pm 2 L_\pm,  \quad [ L_+, L_- ] = L_0. 
\label{8}\end{equation}
For this triplet, 
\begin{equation} 
L_0^*=L_0,\quad  L_+^*= - L_-
\label{8.1}\end{equation}
hold, where $L_+$ and $L_-$ are neither self-adjoint nor skew-adjoint.
Conversely, from the triplet $(L_0,L_+,L_-)$ satisfying the conditions 
(\ref{8}) and (\ref{8.1}), a unitary representation $(E_0,E_+,E_-)$ 
of the algebra $\su(1,1)$ can be constructed by 
\begin{equation}
E_0 = L_+ + L_-, \quad E_\pm = \pm \frac{i}{2} (L_0 \mp L_+ \pm L_-).
\label{9}\end{equation}
The {\it Casimir operator} 
is useful for the analysis of the representation. In case of
the algebra $\su(1,1)$, it is given by 
\begin{equation}
C := E_0^2 + 2( E_+ E_- + E_- E_+)=
L_0^2 + 2 (L_+ L_- + L\lineskip .5em_- L_+).
\label{10}\end{equation}
For the general definition, see p.130-131 of Perelomov \cite{Pe} 
or p.45 of Howe and Tan \cite{HT}.
The relation (\ref{10}) can be written in another form 
\begin{equation}
C= L_0^2 - 2 L_0 + 4 L_+ L_- 
\label{10.1}\end{equation}
by using (\ref{8}). From (\ref{6}) and (\ref{8}), the Casimir operator $C$
is commutative with $E_0,E_+,E_-, L_0,L_+$ and $L_-$. From the Schur's lemma,
in any irreducible representation,
the Casimir operator $C$ is constant.
\begin{lem}
Non-trivial irreducible unitary representations of $\su(1,1)$
are classified into the following three cases:
\begin{eqnarray*}
\lo \hbox{Case 1:} \quad& \dim L_+ = 0 \hbox{ and } \dim L_- = 1 \\
\lo \hbox{Case 2:} \quad& \dim L_+ = 1 \hbox{ and } \dim L_- = 0 \\
\lo \hbox{Case 3:} \quad& \dim L_+ = 0 \hbox{ and } \dim L_- = 0.
\end{eqnarray*}
\end{lem}
Case 2 is reduced to Case 1, by exchanging $L_-$ for $L_+$ 
and by changing the sign of $L_0$. We will not treat Case 3 in this
paper. 
Thus, only Case 1 will be discussed. 

\begin{pf}
The irreducibility requires that 
the dimensions of the kernels of $L_-$ and $L_+$ 
are not more than one. Moreover, if the dimensions of both kernels are one, 
then the representation should be finite-dimensional. However, 
this circumstance is forbidden by the unitarity of the representation.  
Now, the lemma follows immediately.
\end{pf}
\begin{lem}
The unit vector $|0 \rangle_N $ belonging to the Kernel of $L_-$
is a eigen vector of $L_0$.
this eigen value $\lambda$ is called the {\it lowest weight}
and specifies the irreducible unitary representation of $\su(1,1)$ uniquely
and satisfies $\lambda \,> 0$.
The equations 
\begin{eqnarray}
\begin{array}{ll}
L_0 | n \rangle_N &= ( \lambda+ 2n)| n \rangle_N \\
L_+ | n \rangle_N &=  \sqrt{(n+1)(\lambda+n)}| n+1 \rangle_N \\
L_-  | n \rangle_N &= - \sqrt{n(\lambda+n-1)} | n-1 \rangle_N 
\end{array}
\label{720.7}
\end{eqnarray}
hold, where we define $| n \rangle_N := \frac{1}{\|(L_+)^n | 0 \rangle_N\|}
(L_+)^n | 0 \rangle_N$.
\end{lem}
\begin{pf}
Because the Casimir operator should be scalar-valued, 
we can show that $| 0 \rangle_N$ 
is the eigenvector of $L_0$, from (\ref{10.1}). 

Let $v_n := (L_+)^n | 0 \rangle_N$.
The commutation relations (\ref{8}) yields
the following relations;
\begin{equation*}
\begin{array}{ll}
L_0 v_n &= ( \lambda+ 2n) v_n  \\
L_+ v_n &= v_{n+1} \\    
L_- v_n &= - n ( \lambda + n -1) v_{n-1} 
\end{array}
\end{equation*}
whence we can confirm that 
the lowest weight $\lambda$, 
with which $| 0 \rangle_N$ is associated,
specifies the representation uniquely. From the above assumptions,
we can confirm that the basis $\{ v_n \}_{n=1}^{\infty}$ is 
complete and orthogonal. From the above relations, 
the Casimir operator $C$ is calculated to be 
the scalar $\lambda(\lambda-2)$. From 
the commutation relations (\ref{8}), we have 
\begin{equation*}
\langle v_n , v_n \rangle = n (\lambda +n -1) \langle v_{n-1}, v_{n-1} 
\rangle .
\end{equation*}
Therefore, 
the equation 
\begin{equation*}
| n \rangle_N = \sqrt{\frac{\Gamma(\lambda)}{n! \Gamma(\lambda+n)}}
v_n
\end{equation*}
holds. Thus, (\ref{720.7}) follows immediately.
The unitarity 
of the representation guarantees $\lambda \,> 0$. (See Theorem 1.1.5
in pp. 96 of Howe and Tan \cite{HT}.)
\end{pf}
In the following 
discussions, ${\cal H}_{\lambda}$ denotes the representation space of
the irreducible unitary representation of $\su(1,1)$ characterized 
by the lowest weight $\lambda$.
We call such a representation (i.e. Case 1)
{\it lowest-weight-type}.
The representation of the Lie group 
$\SU(1,1)$ 
can no more be constructed 
unless $\lambda$ is an integer than 
the representation of the Lie group 
$\SO(3)$ unless the total momentum is integer.
(For more detail, see Remark \ref{719.1}.)
Especially, when the lowest weight $\lambda$ is an integer, 
the representation of the Lie group 
$\SU(1,1)$ is well-known as the {\it discrete series}
\cite{HT,Kn}. 

\begin{defi}\rm
The operator $N:= \frac{1}{2}(L_0 - \lambda)$ called 
{\it $\su(1,1)$-number operator} because of (\ref{720.7}).
The bounded operator $a: =  \frac{1}{2}L_+^{-1}(L_0- \lambda)
= L_+^{-1}N $ is called the {\it $\su(1,1)$-annihilation operator}.
Its definition is well-defined because 
the vector $N | n \rangle_N$ 
belongs to the range of $L_+$ for any $n$ 
and the kernel of $L_+$ is $\{0\}$.
The {\it $\su(1,1)$-creation operator} is defined 
by the adjoint $a^*$  of $a$.
\end{defi}
The equations
\begin{eqnarray}
a | n \rangle_N &= \sqrt{\frac{n}{n+\lambda-1}} |n -1\rangle_N ,
\label{10.4} \\
 a^* |n\rangle_N &= \sqrt{\frac{n+1}{n+\lambda}} ~|n+1\rangle_N
\label{c2}\end{eqnarray} 
hold, where we mean that $a| 0 \rangle_N= 0$ by (\ref{10.4})  
in the exceptional case where $\lambda=1, n=0$, as a convention. From 
(\ref{10.4}), the commutation relation $[a,N]=a$ is derived. From 
(\ref{10.4}) and (\ref{c2}), 
we have 
\begin{equation}
\eqalign{
a^* a = (N + \lambda -1)^{-1} N , \quad 
a a^* = (N + \lambda)^{-1} ( N + 1 ) , \\
[ a, a^*] = (\lambda-1)( N + \lambda)^{-1}(N + \lambda -1 )^{-1}
}
\label{10.6}
\end{equation}
for $ \lambda \neq 1$, and  
\begin{equation}
a a^*= I ,\quad a^*a = I - | 0 \rangle_N ~_N\langle 0 | ,
\quad
[ a, a^*] = | 0 \rangle_N ~_N\langle 0 |
\label{10.7}
\end{equation}
instead of (\ref{10.6}) for $\lambda=1$.
Next, we will construct the {\it $\su(1,1)$-coherent state} as follows;
\begin{defi}\rm
Introduce the unitary operator  
$U(\xi ):= \exp \left( \xi L_+ - \xi^* L_+^* \right)$
for a complex number $\xi$, according to Perelomov \cite{Pe}. 
For the complex number $\zeta$ such that $| \zeta | \,< 1$, 
we define the {\it $\su(1,1)$-coherent state} $| \zeta \rangle_a$ 
of the algebra $\su(1,1)$ by 
\begin{eqnarray} 
 | \zeta \rangle_a
 : = U\left(\frac{1}{2}e^{i\arg \zeta}\ln\frac{1+|\zeta|}{1-|\zeta|}
\right)~| 0 \rangle_N \label{20-1}.
\end{eqnarray}
\end{defi}
Squeezed states are characterized as $\su(1,1)$-coherent states,
as we discuss in section \ref{s43}.
\begin{lem}
The $\su(1,1)$-coherent state $| \zeta \rangle_a$ 
is an eigenvector of $a$, i.e.
the equation
\begin{eqnarray} 
a | \zeta \rangle_a= \zeta | \zeta \rangle_a \label{23}
\end{eqnarray}
holds.
\end{lem}
\begin{pf} From the definition, 
we have
\begin{eqnarray*}
| \zeta \rangle_a &= \exp(\zeta L_+) \exp 
\left( \frac{1}{2} \ln (1-|\zeta|^2) ~L_0 \right) 
\exp(\zeta^* L_-) ~| 0 \rangle_N  \\
&= (1-|\zeta|^2)^{\lambda/2}\exp(\zeta L_+)~| 0 \rangle_N ,
\end{eqnarray*}
where see pp.73-74 of Perelomov \cite{Pe} for the derivation 
of the first equation. 
Because we can show that $[a, L_+]= I $, we obtain 
the commutation relation 
$\left[ a  ,\exp(\zeta L_+) \right] = \zeta \exp(\zeta L_+)$. 
Moreover, from the relation
$\exp\left( \frac{1}{2} \ln (1-|\zeta|^2) 
~L_0\right) \exp(\zeta^* L_-) ~
| 0 \rangle_N
= (1-|\zeta|^2)^{\lambda/2} |0\rangle_N$,  
we have 
\begin{equation}
a | \zeta \rangle_a
 = \exp(\zeta L_+)~ a ~| 0 \rangle_N + \zeta \exp(\zeta L_+) 
| 0 \rangle_N 
= \zeta | \zeta \rangle_a.
\end{equation}
Therefore, the coherent states of the algebra $\su(1,1)$ are 
characterized as the eigenvectors of the $\su(1,1)$-annihilation 
operator $a$. 
\end{pf}
\begin{lem}\label{720.1}
When $1 \,> \lambda \,> 0$, $a^*$ and $a$ are not subnormal.
When $\lambda \ge 1$, $a^*$ is subnormal and $a$ is not subnormal.
In this case, $a^*$'s POVM is given by 
$(\lambda-1)| \zeta^* \rangle_a ~_a\langle \zeta^*| \mu (\,d\zeta)$,
where we define $\mu (\,d\zeta) 
:= \frac{\,d^2\zeta}{\pi (1-|\zeta |^2)^2}$.
\end{lem}
\begin{pf}
When $\lambda \,> 0$, it is shown that 
$a$ is not subnormal, from Lemma \ref{C13} and 
the fact that it has eigenvectors.
When $\lambda \,< 1$, 
it is shown that $a^*$ is not subnormal, from Lemma \ref{H7} and the fact that
$[a,a^*] \ge 0$ does not hold. (See (\ref{10.6}).)

Moreover, when $\lambda \,> 1$, we can construct 
 the resolution of the identity by the 
system of the coherent states:
\begin{equation}
(\lambda-1)
\int_{D} |\zeta\rangle_a ~_a\langle \zeta| \mu (\,d\zeta) = I
\label{40} 
\end{equation}
where $D$ denotes the unit disk $\{ z \in \complex |~|z| \,<
1\}$. From 
this resolution of the identity and Lemma \ref{H9},
when $\lambda \,> 1$, we can show that $a^*$ is a subnormal operator.
On the other hand, when $\lambda \le 1$, the integral in (\ref{40}) 
diverges. However, the equations (\ref{10.7}) imply that
$a^*$ is isometric when $\lambda = 1$. 
Then, $a^*$ is subnormal even when $\lambda = 1$. 
\end{pf}
\begin{defi}\rm
We formally define the operator 
\begin{equation*}
A := -i (a + 1) ( a - 1 )^{-1}.
\end{equation*}
Since this operator is unbounded,
we need more attention in this definition.
First, define the unbounded operator $\tilde{A}$ 
by a linear fractional transform (M\"{o}bius transform) of $a$, as
$\tilde{A} := -i (a + 1) ( a - 1 )^{-1}$, 
where the domain $\odom(\tilde{A})$ of $\tilde{A}$ is defined by 
$ \langle \{  | n \rangle_N \}_{n=0}^{\infty} 
\rangle$.
The domain of $\tilde{A}^*$ is dense in ${\cal H}_{\lambda}$, 
as will be shown in the last part of Remark \ref{720.4}. Therefore, 
$\tilde{A}$ is closable and we can define the operator $A$ by 
$A:=\overline{\tilde{A}}= \tilde{A}^{**}$.
(See Reed and Simon \cite{RS}.)
\end{defi}
It is shown that $|\zeta \rangle_{a}\in \odom(A)$ 
in the last part of Remark \ref{720.4}. 
Hence we have 
$A| \zeta\rangle_a = -i \frac{\zeta+1}{\zeta-1}| \zeta \rangle_a$. 
By defining 
$| \eta \rangle_A:= \left| \frac{\eta-i}{\eta+i} \right\rangle_a$, 
we can show that 
\begin{equation}
A| \eta \rangle_A= \eta | \eta \rangle_A \label{19.1}
\end{equation}
holds.
(Formally, the operator $a$ is the Cayley transform of $A$, 
with an appropriate discussion on its domain.) 
\begin{lem}
We have another expression of $A$:
\begin{equation} 
A = \frac{1}{2} E_+^{-1} (E_0 - \lambda).
\label{27}\end{equation}
\end{lem}
\begin{pf} From the relations 
$[a, L_+]= I $, (\ref{8}), (\ref{9}) 
and the definition of $a$, we can show that the relations 
\begin{eqnarray*}
2 (E_0 - \lambda) (a-1) L_+ 
&=& (L_+ + L_- - \lambda) \bigl(L_0 - (\lambda -2) - 2 L_+\bigr) \\ 
&=& (L_0 - L_+ + L_-) ( -\lambda + L_0 + 2 + 2 L_+)\\
&=& -4i E_+ (a+1) L_+  \\
(E_0 - \lambda)(a-1) | 0 \rangle_N
&= & - (E_0 - \lambda) | 0 \rangle_N
=(L_0 - L_+ ) | 0 \rangle_N \\
&=& -2i E_+ | 0 \rangle_N
= -2i E_+ (a+1) | 0 \rangle_N 
\end{eqnarray*}
hold on $\odom(\tilde{A})$.
Hence, on $\odom(\tilde{A})$, we have 
\begin{equation} 
(E_0 - \lambda) ( a - 1 ) = -2i E_+ ( a + 1 ).
\label{25}\end{equation}
By using (\ref{25}), we obtain (\ref{27}).
\end{pf}From (\ref{27}), we have 
\begin{equation*}
[A,A^*]= -(\lambda-1)E_+^{-2}
\end{equation*}
formally, and 
\begin{equation}
A^* A - A A^* = (\lambda-1) \left( E_+^{-1}\right)^* E_+^{-1}
\quad 
\cases{
\hbox{ on } \odom(A A^*) & for $\lambda \ge 1$ \\
\hbox{ on } \odom(A^* A) & for $0 \,< \lambda \,< 1$ 
} \label{11.1}
\end{equation}
in more precise form. 
(The proof of this relation will be given in Remark \ref{720.4}.)

\begin{lem}\label{720.2}
When $1 \,> \lambda \,> 0$, $A$ and $A^*$ are not subnormal.
When $\lambda \ge 1$, $A^*$ is subnormal and $A$ is not subnormal.
In this case,
$A^*$'s POVM is given by $(\lambda-1)| \eta^* \rangle_A ~_A\langle
\eta^* | \nu( \,d \eta )$, where 
we define $\nu(\,d\eta) 
:= \frac{\,d^2\eta}{4\pi({\rm Im}~\eta)^2}$.
\end{lem}
\begin{pf}
For $\lambda \,> 0$, 
from Lemma \ref{C13} and the fact that 
the operator $A$ has eigenvectors, 
it is shown that $A$ is not subnormal.
When $\lambda \,< 1$, $A^*$ is not 
subnormal because the relation (\ref{10.6}) shows that 
the condition $AA^* \ge A^* A $ is not satisfied. 
Moreover, 
in a similar manner to the above discussion, the resolution of 
the identity by the eigenvectors of $A$
\begin{equation*}
(\lambda-1)
\int_{\rm H} |\eta\rangle_A ~_A\langle \eta| \nu (\,d\eta) = I
\end{equation*}
holds when $\lambda \,> 1$. Hence, when $\lambda \,> 1$, we can show 
that $A^*$ is subnormal from Lemma \ref{H9}.
When $\lambda \le 1$, the integral in (\ref{40}) 
diverges. 
However, as will be proved in the last part of Remark \ref{720.4}, 
the operator $A^*$ is maximal symmetric when $\lambda =1$. From Lemma 
\ref{H10}, $A^*$ is subnormal even when $\lambda = 1$. 
\end{pf}
\begin{remark}\label{719.1}\rm
[Relation to unitary representations of $\SU(1,1)$]
In the following, we discuss Definition \ref{719.2} from the viewpoint of
a unitary representation of the group $\SU(1,1)$.
Any element $g$ in the group $\SU(1,1)$
is specified by two complex numbers 
$\nu(g)=\nu_1(g) + \nu_2(g) i , \mu(g)=\mu_1(g) + \mu_2(g) i$
satisfying 
$|\nu(g)|^2- |\mu(g)|^2 = 1$
as
\begin{eqnarray*}
g = \left(
\begin{array}{cc}
\mu^*(g) & \nu(g) \\
\nu^*(g) & \mu(g)
\end{array}
\right) 
=
\left(
\begin{array}{cc}
\mu_1(g) - \mu_2(g) i & \nu_1(g) + \nu_2(g) i \\
\nu_1(g) - \nu_2(g) i & \mu_1(g) + \mu_2(g) i
\end{array}
\right) .
\end{eqnarray*}
The group $\SU(1,1)$ is isomorphic to the group $\SL(2,\real)$
by the map
\begin{eqnarray}
\fl
j: 
\left(
\begin{array}{cc}
\mu_1(g) - \mu_2(g) i & \nu_1(g) + \nu_2(g) i \\
\nu_1(g) - \nu_2(g) i & \mu_1(g) + \mu_2(g) i
\end{array}
\right)
\mapsto 
\left(
\begin{array}{cc}
\mu_1(g) + \nu_1(g) & - \mu_2(g) - \nu_2(g) \\
\mu_2(g) - \nu_2(g) & \mu_1(g) - \nu_1(g)
\end{array}
\right)
. \label{2.11}
\end{eqnarray}
The (Lie) algebra $\su(1,1)$ associated with $\SU(1,1)$ is written
by
\begin{eqnarray*}
\fl
\su(1,1)
= 
\left\{\left.
\left(
\begin{array}{cc}
a_{1,1} & a_{1,2} \\
a_{2,1} & a_{2,2}
\end{array}
\right)
\right|
\left(
\begin{array}{cc}
a_{1,1}^* & - a_{2,1}^* \\
-a_{1,2}^* & a_{2,2}^*
\end{array}
\right)
=
-
\left(
\begin{array}{cc}
a_{1,1} & a_{1,2} \\
a_{2,1} & a_{2,2}
\end{array}
\right)
, a_{1,1} + a_{2,2} = 0 
\right\} .
\end{eqnarray*}
The vector space $\su(1,1)$ has the following basis $e_0 ,e_+ , e_-$ as
\begin{eqnarray*}
e_0 = 
\left(
\begin{array}{cc}
0 & 1 \\
1 & 0
\end{array}
\right),
e_+ =
\frac{i}{2}
\left(
\begin{array}{cc}
1 & -1 \\
1 & -1
\end{array}
\right),
e_- =
\frac{i}{2}
\left(
\begin{array}{cc}
-1 & -1 \\
1 & 1
\end{array}
\right).
\end{eqnarray*}
Thus,
from the isomorphism (\ref{2.11}), we can naturally define
the isomorphism $j_*$ from the algebra $\su(1,1)$ to the algebra 
$\sl(2,\real)$.
Then, the image $j_*(e_0) , j_*(e_-), j_*(e_+)$ 
of the basis is written as
\begin{eqnarray}
j_*(e_0) = 
\left(
\begin{array}{cc}
1 & 0 \\
0 & -1
\end{array}
\right),
j_*(e_+) =
\left(
\begin{array}{cc}
0 & 1 \\
0 & 0
\end{array}
\right),
j_*(e_-) =
\left(
\begin{array}{cc}
0 & 0 \\
1 & 0
\end{array}
\right). \label{6.1.1}
\end{eqnarray}
The basis $e_0, e_- ,e_+$ satisfies the following commutation relation:
\begin{eqnarray}
[ e_0, e_\pm ] = \pm 2 e_\pm,  \quad [ e_+, e_- ] = e_0. 
\label{6.1}
\end{eqnarray}
A map $V$
from a group $G$ to the set of unitary operators on a Hilbert space ${\cal H}$
is called a {\it unitary representation} of the group $G$ on ${\cal H}$
if 
\begin{eqnarray*}
V(g_1 g_2) = V( g_1 ) V(g_2), \quad \forall g_1, g_2 \in G .
\end{eqnarray*}
Let $\mathfrak{g}$ be the Lie algebra associated with a Lie group
$G$. From a unitary representation of the group $G$ on a Hilbert 
space ${\cal H}$,
we can naturally define the map $V_*$ from the Lie algebra $\mathfrak{g}$
to the set of skew-adjoint operators on ${\cal H}$, by 
\begin{eqnarray*}
V_* ( X) := 
\frac{ \,d V ( \exp (t X) )}{\,d t}|_{t=0}.
\end{eqnarray*}
It satisfies that
$V_* ( [ X,Y] )=[ V_*(X), V_*(Y) ]$.
Then, a linear map $f$ from a Lie algebra $\mathfrak{g}$ to
the the set of skew-adjoint operators on a Hilbert space ${\cal H}$
is called a {\it unitary representation} of the Lie algebra $\mathfrak{g}$
on ${\cal H}$ if
\begin{eqnarray*}
[ f(X) , f(Y) ] = f( [ X,Y] ), \quad
\forall X ,Y \in \mathfrak{g}.
\end{eqnarray*}
We can construct the unitary representation $V$ of the universal covering
group\footnote{A group is called a {\it universal covering} group
if it is connected and if its homotopy group is trivial.}
$\hat{G}$ associated with a Lie algebra $\mathfrak{g}$ from 
a unitary representation $f$ of $\mathfrak{g}$, by 
\begin{eqnarray*}
V( \exp X) := \exp f( X), \quad \forall X \in \mathfrak{g}.
\end{eqnarray*}
Since any element of the (Lie) algebra $\su(1,1)$ is described by a linear sum
of bases $e_0, e_+,e_-$,
we can uniquely construct the unitary representation of the algebra $\su(1,1)$
from a triplet $(E_0,E_+,E_-)$ of skew-adjoint operators
satisfying (\ref{6}).
Thus, we can regard 
the triplet $(E_0,E_+,E_-)$ satisfying (\ref{6})
as the unitary representation of the algebra $\su(1,1)$.
\end{remark}

\begin{remark}\rm 
[Spectrums of $a,a^*,A,A^*$]
The point spectrum $\sigma_p(a)$ 
is the open unit disk $D$
the continuous spectrum $\sigma_c(a)$ is 
the unit circle $S:= \{ z \in \complex ||z| = 1\}$ 
and the residual spectra is the empty set.
Moreover, from Lemma \ref{HA1}, 
the point spectrum $\sigma_p(a^*)$, 
the continuous spectrum $\sigma_c(a^*)$ 
and the residual spectrum $\sigma_r(a^*)$ of $a^*$ are 
the empty set, $S$ and $D$, respectively.

It is shown that the point spectrum $\sigma_p(A)$ of $A$ is 
the upper-half plane $H$, 
the continuous spectrum $\sigma_c(A)$ is the real axis $\real $ 
and the residual 
spectrum $\sigma_r(A)$ is the empty set.  On the other hand, 
from Lemma \ref{HA1}, the point spectrum $\sigma_p(A^*)$, 
the continuous spectrum $\sigma_c(A^*)$ 
and the residual spectrum $\sigma_r(A^*)$ of $A^*$ 
are the empty set, $\real$ and $H$, respectively.
\end{remark}

\begin{remark}\label{720.5}
\rm
[Action of the group to operators $a,A$]
First, we discuss the action to a operator $a$.
We let $\pi_{\widehat{\SU(1,1)}}$ be the projection from 
the universal covering group $\widehat{\SU (1,1)}$ 
to the group $\SU(1,1)$,
and let $U$ be the maximal Cartan subgroup of $\widehat{\SU (1,1)}$ 
i.e. the 1-parameter subgroup 
generated by $i L_0$.

The homogeneous space $\widehat{\SU(1,1)}/ U$ is isomorphic to the open unit 
disk $D$
in the sense that an element $g$ of the group $\widehat{\SU(1,1)}$ acts on the
open unit disk $D$ as
\begin{eqnarray*}
\zeta \mapsto \frac{\mu^*  \zeta + \nu  }
{\nu^* \zeta + \mu},
\quad \zeta \in D,
\end{eqnarray*}
where we simply use the notations $\mu$ and $\nu$ instead of    
the complex numbers $\mu \circ \pi_{\widehat{\SU(1,1)}} ( g )$
and $\nu \circ \pi_{\widehat{\SU(1,1)}} ( g )$ 
with the functions $\mu$ and $\nu$ defined 
at the beginning of Remark \ref{719.1},
respectively.
We let $V$ be the representation of the group $\widehat{\SU(1,1)}$, defined by
this representation of $\su(1,1)$.
Then, we have 
\begin{eqnarray}
\fl
V(g) | \zeta \rangle_a ~_a\langle \zeta | V(g)^* = 
\left| \frac{\mu^* \zeta + \nu }{\nu^* \zeta + \mu} 
\right\rangle_a
~_a\left\langle  \frac{\mu^* \zeta + \nu }{\nu^* \zeta + \mu} \right| ,
\quad g \in \widehat{\SU (1,1)} , \zeta \in \complex .
\label{724.1}
\end{eqnarray}
Thus, for any element $g \in \widehat{\SU (1,1)}$ and any complex number $\zeta$,
there exists a real number $\theta (g, \zeta )$
such that
\begin{eqnarray}
V(g) | \zeta \rangle_a = e^{i  \theta (g, \zeta )} 
\left| \frac{\mu^* \zeta + \nu }{\nu^* \zeta + \mu} \right\rangle_a .
\label{22}
\end{eqnarray}
the equations (\ref{23}) and (\ref{22}) implies that
\begin{eqnarray}
V(g)^* a  V(g)| \zeta \rangle_a = 
\frac{\mu^* \zeta + \nu }{\nu^* \zeta + \mu}
| \zeta \rangle_a. \label{22.1}
\end{eqnarray}
Since the subspace $\langle \{ | \zeta \rangle_a \} \rangle$\footnote{$\langle X \rangle $ denotes the vector space whose 
elements are finite linear sums of a set $X$.}
is dense,
we obtain 
\begin{eqnarray*}
V(g)^* a  V(g) = (\mu^* a + \nu )(\nu^* a + \mu)^{-1},
\end{eqnarray*}
where we can define
the bounded operator $(\nu^* a + \mu)^{-1}$
by $(\nu^* a + \mu)^{-1}:=
\frac{1}{\mu}
\sum_{n=1}^\infty \left(- \frac{\nu^*}{\mu} a \right)^n $
because the norm of
the operator $- \frac{\nu^* }{\mu}a $
is less than 1.

Next, we consider the action to the operator $A$.
Similarly to (\ref{22.1}), we have
\begin{eqnarray*}
V(g) A V(g)^*| \eta \rangle_A =
\frac{  (\mu_1 + \nu_1) \eta -\mu_2 - \nu_2 }{ (\mu_2 - \nu_2) \eta 
+\mu_1 - \nu_1 }
| \eta \rangle_A,
\end{eqnarray*}
where we simplifies $\mu_i \circ \pi_{\widehat{\SU(1,1)}}( g)$
and $\nu_i \circ \pi_{\widehat{\SU(1,1)}}( g)$
as $\mu_i$ and $\nu_i$, respectively.
\end{remark}

\section{Concrete representations of $\su(1,1)$}
\subsection{Representation associated 
with irreducible unitary representation 
of affine group}\label{s42}
Next, we will construct lowest-weight-type 
irreducible unitary representations of the algebra $\su(1,1)$ 
from an irreducible unitary representation of the affine group 
($ax+b$ group)
generated by $E_+$ and $E_0$. The representation which will be 
constructed in this section is closely related to the continuous 
wavelet transformation \cite{DaSIAM,sita}. 
In this representation, the pair $A$ and $| \eta \rangle_A$
plays a more important role than the pair
$a$ and $|\zeta\rangle_a$.
According to Aslaksen and Klauder \cite{AK}, there is not 
any irreducible representation of the affine group 
but the representations equivalent unitarily 
to the following representation on $L^2(\real^+)$ or $L^2(\real^-)$;
\begin{eqnarray}
E_0= i ( PQ+ QP ) ,\quad E_+ = iQ ,
\label{50}
\end{eqnarray}
where $E_0$ and $E_+$ are shown to be skew adjoint.
In this representation, the vector
$\sqrt{\frac{ (2 \im \eta)^{2k+1}}{\Gamma(2k+1)}}
x^k e^{i \eta x}$
is called the {\it affine coherent state}\footnote{The 
Fourier transform of this affine coherent state 
is equivalent to the Cauchy wavelet in signal processing, whose 
basic wavelet function is $\frac{({\rm const.})}{(t \pm i)^{k+1}}$.},
and it is obtained 
by operating the affine group on the {\it affine vacuum state}
$\sqrt{\frac{ 2^{2k+1}}{\Gamma(2k+1)}}
x^k e^{\mp x}$.
In the following, we will construct an irreducible unitary 
representation of the algebra $\su(1,1)$ from the above type 
of unitary representation of the affine group, and 
will discuss how to interpret the affine coherent states 
in terms of the unitary representation of the algebra 
$\su(1,1)$. Therefore, in addition to the two generators in 
(\ref{50}), we should introduce the representation of 
another additional generator $E_-$.  By choosing 
\begin{eqnarray}
\tilde{E}_{-,k} := -i( P Q P+ k^2 Q^{-1}) \quad (k \,> -1/2) \label{51}
\end{eqnarray}
for this additional generator, we can construct an irreducible 
unitary representation where the triplet $E_0, E_+$ and $E_-$ satisfies 
the commutation relations (\ref{6}).
However, we should be careful about the domain of $\tilde{E}_{-,k}$, 
as follows; first, define the dense subspace $\odom(\tilde{E}_{-,k})$ 
of $L^2(\real^+)$ by 
\begin{eqnarray*}
\fl \odom(\tilde{E}_{-,k}) \\
\fl :=\left\{ f(x)= x^k f_0 (x) \in L^2(\real^+) \cap
C^1(\real^+)
 \left| 
\begin{array}{l}
(2k+1)x^k f_0'(x) + x^{k+1} f_0''(x) \in L^2(\real^+),\\
\displaystyle \limsup_{s \to 0} f_0 (s) \,< \infty , \quad
x^k f_0 (x) \to 0 \hbox{ as } x \to \infty
\end{array}
\right. \right\} .
\end{eqnarray*}
Then $\tilde{E}_{-,k}$ is an operator defined on $\odom(\tilde{E}_{-,k})$.
We need attention to the domain when $-\frac{1}{2} \,< k \,< \frac{1}{2}$.
\begin{lem}
The operator $\tilde{E}_{-,k}$ has 
the skew-adjoint extension, uniquely.
\end{lem}
In the following, its skew-adjoint extension is written by
$E_{-,k}$.

\begin{pf}
It is confirmed that $i \tilde{E}_{-,k} = P Q P+ k^2 Q^{-1}$ is 
a symmetric operator on $\odom(\tilde{E}_{-,k})$, from the fact that 
the difference 
\begin{eqnarray*}
\fl \int_s^t 
\left( \left( P Q P + k^2 Q^{-1} \right) f\right)^* (x) g(x) \,d x 
-
\int_s^t (f (x) )^*
\left( \left( P Q P + k^2 Q^{-1} \right)g\right) (x) \,d x \\
\lo=
\left[
x (f' (x))^* g(x) - x g' (x) 
(f (x) )^*
\right]_s^t \\
\lo= 
t (f' (t))^* g(t) - t g' (t) (f (t) )^* \\
-\left(
s g(s) \left( (f' (s))^* - \frac{k}{s} (f (s))^* \right)
- s ( f (s))^* \left( g' (s) - \frac{k}{s}g(s) \right)\right) \\
\lo=
t ( f' (t))^* g(t) - t g' (t) (f (t) )^*
-\left(
s g(s) s^k ( f_0'(s) )^*
- s (f (s))^* s^k g_0' (s) 
\right) 
\end{eqnarray*}
tends to zero as $s \to 0 ,t \to \infty$.
Since $i \tilde{E}_{-,k}$ is semi-bounded, 
Friedrich extension theorem guarantees that 
there uniquely exists the self-adjoint extension of $i \tilde{E}_{-,k}$.
(See pp. 177 of Reed and Simon \cite{RS2}.)
Now, the proof is complete.
\end{pf}
By letting 
$L_{+,k},L_{-,k} ,L_{0,k},\tilde{A}_k,A_k, N_k, | n \rangle_N^k$ 
and $|\eta \rangle_A^k$ be 
$L_+,L_- ,L_0,\tilde{A},A,N,| n \rangle_N $
and $|\eta \rangle_A$ in this representation, respectively, we have 
\begin{eqnarray*}
L_{+,k} =
\frac{1}{2}\left(
i(P Q + Q P)- Q + P Q P + k^2 Q^{-1} \right), \\
L_{-,k} =
\frac{1}{2}\left(
i(P Q + Q P)+Q - P Q P - k^2 Q^{-1} \right),  \\
L_{0,k} = \left( P Q P+ k^2 Q^{-1}+ Q\right), \quad
\tilde{A}_k= P + i k Q^{-1} , \\
N_k= \frac{1}{2} \left( P Q P+ k^2 Q^{-1}+ Q -1 - 2k\right), \\
| n \rangle_N^k (x) =
\sqrt{\frac{2^{2k+1} n!}{\Gamma(n+2k+1)}}
e^{-x}x^k S_n^{2k}(2x) , \\
|\eta \rangle_A^k (x)= \sqrt{\frac{ (2 \im \eta)^{2k+1}}{\Gamma(2k+1)}}
x^k e^{i \eta x} , 
\end{eqnarray*}
when $S_n^l(x)$ is the Sonine Polynomial (or the associated 
Laguerre polynomial) defined by\footnote{Sometimes another 
definition with $n+l$ instead of $l$ is used.} 
\begin{eqnarray*}
S_n^l(x):=
\sum_{m=0}^n 
\frac{(-1)^m}{(n-m)!}\frac{\Gamma(n+l+1) x^m}{\Gamma(m+l+1)m!} . 
\end{eqnarray*}
$|\eta\rangle_A^k(x)$ is the affine coherent state.
Moreover, in this representation, the minimum eigenvalue of $L_{0,k}$ is 
$\lambda= 2k+1$, and the Casimir operator is $4k^2-1$. 
Thus, we have the following theorem.
\begin{thm}\label{720.3}
The representations of lowest-weight-type (defined  
in section \ref{s3}) in general can be concretely constructed 
by (\ref{50}) and (\ref{51}) on $L^2(\real^+)$
in the correspondence $\lambda= 2k+1$.
\end{thm}
\begin{remark}\label{720.4}\rm
[Domains of $A_k,A_k^*$]
In the following, we will show the properties of $\tilde{A}_k$ 
in order to show the properties of $\tilde{A}$ 
in the representations of lowest-weight-type.
Since the domain of $\tilde{A}_k$ is 
$\langle \{ | n \rangle \}_{n=0}^{\infty} \rangle$ 
and $\tilde{A}_k= P + i k Q^{-1}$, 
the  relation 
\begin{eqnarray}
\fl \odom(\tilde{A}_k^*) \cap C^1(\real^+) =
\left\{
x^{-k} f(x) \in L^2(\real^+) \cap C^1(\real^+) \left|
\begin{array}{l}
x^{-k} f'(x) \in L^2(\real^+) \\
f(s) \to 0 \hbox{ as } s \to 0,
\end{array} \right. \right\} \label{11.2} 
\end{eqnarray}
is derived, 
and hence we can show that 
$\odom(\tilde{A}_k^*)$ is dense in $L^2(\real^+)$. 
Thus $\tilde{A}_k$ is shown to be a closable operator.
Since $A_k= \overline{\tilde{A}_k}$,
the relation 
\begin{eqnarray}
\fl \odom( A_k ) \cap C^1(\real^+)
= \left\{
x^{k} f(x) \in L^2(\real^+) \cap C^1(\real^+) \left|
\begin{array}{l}
x^{k} f'(x)  \in L^2(\real^+), \\
\limsup_{s \to 0} f(s) \,< \infty  
\end{array}
\right. \right\}. \label{11.3}
\end{eqnarray}
is confirmed. Note that 
$\overline{X}=X^{**}$ and $X^*=\overline{X}^*$ hold for
a densely defined linear operator $X$, and 
that $\limsup_{x \to \infty} f_0(x)=0$ 
for $ k \,>- \frac{1}{2} $ when 
$x^k f_0(x) \in L^2 (\real^+)$.  From 
(\ref{11.3}), we can show that $| \zeta \rangle_a \in \odom(A_k)$. 
Thus, the subspaces $\odom(A_k^*) \cap C^1(\real^+)$ and $\odom(A_k)
\cap C^1(\real^+)$ are 
the cores \footnote{A subspace of the domain $\odom(X)$ 
of a closed operator $X$ is called a {\it core} of $X$
if it is dense in $\odom(X)$ with respect to the graph norm 
of the operator $X$.} 
of $A_k^*$ and $A_k$.
When $- \frac{1}{2} \,< k \,< \frac{1}{2}$, 
the domain of $A_k$ is larger than the domain of $A_{-k}^*$, 
though $A_k$ and $A_{-k}^*$ are the same formally, i.e.
$A_{-k}^* \subsetneqq A_k$.
In the special case where $\lambda =1$ (i.e. where $k=0$), 
$A^*_0$ is symmetric.
Since $A_0= A_0^{**}$ has no spectrum in the lower half plane,
$A_0^*$'s deficiency indices are $(1,0)$.
(For the definition of deficiency indices, 
see p.138 of Reed and Simon \cite{RS2} or p.360 of Rudin \cite{Ru}.)
Therefore, the operator $A_0^*$ is a maximal symmetric operator.

These relations 
$\odom(A_k^*) \subset \odom(A_k)$ and $\odom(A_k^*)  \subset \odom(E_+^{-1})$ are shown 
in the cases where $\lambda \,< 1 ~( k \,> 0)$, 
only the relation 
$\odom(A_k^*) \subset \odom(A_k)$ is shown when $\lambda = 1 ~(k=0)$, 
and these relations
$\odom(A_k) \subset \odom(A_k^*)$ and 
$\odom(A_k) \subset \odom(E_+^{-1})$ are shown when 
$0 \,< \lambda \,< 1 ~( - \frac{1}{2} \,< k \,< 0)$. From Theorem \ref{720.4},
these discussions and (\ref{27}), we obtain (\ref{11.1}).
\end{remark}

\subsection{Representation associated with squeezed states}\label{s43} 
Next, we will discuss the following representation 
of the algebra $\su(1,1)$ on the Hilbert space $L^2(\real)$; 
let 
\begin{equation}
E_0= \frac{i}{2} (P Q+Q P) , \quad E_+= \frac{i}{2} Q^2, 
\quad E_-= -\frac{i}{2} P^2 , \label{99.4}
\end{equation}
then we have 
\begin{equation}
L_0= n_b + \frac{1}{2}, \quad
L_+= -\frac{1}{2}(a_b^*)^2, \quad
L_-= \frac{1}{2}(a_b)^2 , \label{2.5}
\end{equation}
where the boson annihilation operator $a_b$ and 
the boson number operator $n_b$ are given by
$a_b= \sqrt{\frac{1}{2}}(Q+iP)$ and $n_b = \frac{1}{2}(Q^2+P^2-1)= a_b^* a_b$.
In this representation, the Casimir operator is the 
scalar $-\frac{3}{4}$. From the fact that the Casimir operator 
is the scalar $\lambda(\lambda-2)$, 
the solutions are $\lambda=1/2,3/2$. 
Under the representation given in (\ref{99.4}), 
$L^2(\real)$ is not irreducible and it is decomposed 
into two irreducible subspaces as:
\begin{equation*}
L^2(\real)
= L^2_{\rm even}(\real) \oplus L^2_{\rm odd}(\real) 
\end{equation*}
where $L^2_{\rm even}(\real)$ is the set 
of square-integrable even functions and 
$L^2_{\rm odd}(\real)$ is the set 
of square-integrable odd functions.
The solution $\lambda=1/2$ corresponds to the subspace 
$L^2_{\rm even}(\real)$ and
the solution $\lambda=3/2$ does to the subspace 
$L^2_{\rm odd}(\real)$.
In the subspace $L^2_{\rm even}(\real)$,
the operators $a,A$ and $N$ are written in the forms  
\begin{eqnarray*} 
a = -(a_b^*)^{-1} a_b  , \quad
A= Q^{-1}P , \quad
N = \frac{1}{2}n_b , \quad
|n \rangle_N= (-1)^n|2n \rangle_{n_b}, \\
 \odom(A) \cap C^1(\real)=
\left\{ f \in L^2_{\rm even}(\real) \cap C^1(\real) \left|
\frac{1}{x}f'(x) \in L^2(\real) \right.\right\} , 
\end{eqnarray*}
where $|n \rangle_{n_b}$ denotes the eigenvector in $L^2(\real)$ 
of the boson number operator $n_b$ associated with the eigenvalue $n$.

\begin{lem}
In the action of $\su(1,1)$ on $L^2_{\rm even}(\real)$,
we have
\begin{eqnarray}
\fl
| 0 ; \mu , \nu \rangle \langle 0 ; \mu , \nu |
=
V(g)| 0 \rangle_a ~_a\langle 0 | V(g)^* 
=
\left|  \frac{\nu}{\mu}  \right\rangle_a 
~_a\left\langle \frac{\nu}{\mu} \right|
=
\left| i \frac{\mu+ \nu}{\mu-\nu} \right\rangle_A
\left._A \left\langle i \frac{\mu+ \nu}{\mu-\nu} \right| \right. 
\label{2.9}.
\end{eqnarray}
\end{lem}
Remark that the squeezed state $| 0 ; \mu , \nu \rangle$
is defined as the unit eigen vector of $b_{\mu,\nu}=
\mu a_b + \nu a_b^*$ 
associated with the eigen value $0$.

\begin{pf}
We need the discussion of Remark \ref{719.1} and \ref{720.5} for the proof.
It is necessary to discuss the action of the group. From
this representation of the algebra $\su(1,1)$,
we can construct the representation of the double-covering group 
$\widetilde{\SU
(1,1)}$ of the group $\SU(1,1)$.
In general, we can construct the representation of $\widetilde{\SU(1,1)}$ 
in the case where $\lambda$ is a half-integer.
Now, we let $\pi_{\widetilde{\SU(1,1)}}$
be the projection from $\widetilde{\SU(1,1)}$ to 
$\SU(1,1)$. From (\ref{99.4}), we have 
\begin{eqnarray*}
e^{tE_0} Q e^{-tE_0} = e^{t} Q ,\quad
e^{tE_+} Q e^{-tE_+} = Q,\quad
e^{tE_-} Q e^{-tE_-} = Q - P t , \\
e^{tE_+} P e^{-tE_+} = P - Q t,\quad
e^{tE_0} P e^{-tE_0} = e^{-t} P,\quad
e^{tE_0} P e^{-tE_0} = P.
\end{eqnarray*}From (\ref{2.11}), (\ref{6.1.1}) and some calculations, 
we have 
\begin{eqnarray*}
V(g) Q V(g)^* = 
(\mu_1 + \nu_1) Q + ( \nu_2 - \mu_2) P, \\
V(g) P V(g)^* = ( \nu_2 + \mu_2)Q +  (\mu_1 - \nu_1 )P , \quad
\forall g \in \widetilde{\SU(1,1)},
\end{eqnarray*}
where the complex numbers $\mu_i \circ \pi_{\widetilde{\SU(1,1)}} 
( g )$
and $\nu_i \circ \pi_{\widetilde{\SU(1,1)}} ( g )$ 
with the functions $\mu_i$ and $\nu_i$ defined at the beginning 
of Remark \ref{719.1} are denoted simply 
by $\mu_i$ and 
$\nu_i$, respectively, 
in a similar manner to the previous section.
Thus, we have
\begin{eqnarray*}
V(g) a_b V(g)^* = \mu a_b + \nu a_b^*, \\
V(g) a_b^* V(g)^* = \nu^* a_b + \mu^* a_b^*, \quad
\forall g \in \widetilde{\SU(1,1)} , 
\end{eqnarray*}
where we simplifies $\mu \circ \pi_{\widetilde{\SU(1,1)}}( g)$
and $\nu \circ \pi_{\widetilde{\SU(1,1)}}( g)$
by $\mu$ and $\nu$, respectively, similarly.

Since $L_- = \frac{1}{2}(a_b)^2$, 
the lowest weight vector $| 0 \rangle_a $ is 
the boson vacuum vector $| 0 ; 1, 0 \rangle$.
The squeezed state $| 0 ; \mu , \nu \rangle$ satisfies 
$(\mu a_b + \nu a_b^* ) | 0 ; \mu , \nu \rangle = 0$.
Assume that $\mu \circ \pi_{\widetilde{\SU(1,1)}}(g)= \mu,
\nu \circ \pi_{\widetilde{\SU(1,1)}}(g)= \nu$. Then we have
$V(g) a_b V(g)^*  | 0 ; \mu , \nu \rangle= 0$.
Hence, we see that the vector $V(g)^* | 0 ; \mu , \nu \rangle$
equals a scalar-times vacuum vector $| 0 ; 1 , 0 \rangle = | 0
\rangle_a$. From these facts and (\ref{724.1}),
we obtain (\ref{2.9}).
\end{pf}From (\ref{2.9}),
we find the correspondence to the characteristic 
equations (\ref{5}) and (\ref{19.2}) of squeezed states 
explained in section \ref{intr}.
Substituting (\ref{2.9}) into (\ref{23}) and (\ref{19.1}), 
we obtain (\ref{19.2}) and (\ref{5}). In the following, 
the vector $| \zeta \rangle_a$ in $L^2_{\rm even}(\real)$ is denoted 
by $| \zeta \rangle_{a,{\rm even}}$. 
The equations (\ref{20-1}) and (\ref{2.9}) implies that
the squeezed state $| 0 ; \mu , \nu \rangle$ equals a scalar-times 
$\exp (-\frac{\xi}{2}(a_b^*)^2 + \frac{\xi^*}{2}(a_b)^2)
| 0 ; 1 , 0 \rangle$
corresponding to Caves's notation \cite{Caves} of squeezed state,
where $\xi:= \frac{1}{2}e^{i {\rm arg} \frac{\nu}{\mu}}\ln
\frac{|\mu|+ | \nu |}{|\mu|- | \nu|}$.

On the other hand, in $L^2_{\rm odd}(\real)$, the operators $a,A$ and $N$ are 
written in the forms 
\begin{eqnarray*}
a = - a_b (a_b^*)^{-1}  , \quad
A= P Q^{-1} , \quad
N = \frac{1}{2}(n_b-1), \quad 
|n \rangle_N = (-1)^n |2n +1\rangle_{n_b}.
\end{eqnarray*}

Next, we will discuss the representation of the algebra $\su(1,1)$ 
in the Hilbert space $L^2(\real^n)= 
\underbrace{L^2 (\real) \otimes \cdots \otimes  L^2 (\real)}_{n} $
, for multi-particle systems.
In this representation, 
\begin{equation}
E_0= \frac{i}{2} \sum_{j=1}^n \left(P_j Q_j + Q_j P_j \right)
,\quad E_+= \frac{i}{2} \sum_{j=1}^n Q_j^2 
, \quad 
E_-= -\frac{i}{2} \sum_{j=1}^n P_j^2 , 
\label{910.1}
\end{equation}
hold, where $Q_j$ and $P_j$ denotes the multiplication operator 
and the $(-i)$-times differential operator, respectively, with 
respect to the $j$-th variable.
Let $L^2_{\rm e}(\real^n)$ be the closure of the linear space generated by 
$\{ |\zeta \rangle_{a,{\rm even}}^{\otimes n}:=
\underbrace{|\zeta \rangle_{a,{\rm even}} \otimes \cdots 
\otimes |\zeta \rangle_{a,{\rm even}}}_{n} \}  $.
Then, the Hilbert space $L^2_{\rm e}(\real^n)$ is 
irreducible under the representation (\ref{910.1}) of the algebra 
$\su(1,1)$, and then we have $L^2_{\rm e}(\real^n)= \{ f \in L^2(\real^n) |
f~ \hbox{is a function of}~\sum_{j=1}^n x_j^2 \}$, 
and then the vector 
$| \zeta \rangle_a$ in this representation on $L^2_{\rm e}(\real^n)$ 
is equivalent to $|\zeta \rangle_{a,{\rm even}}^{\otimes n}$.

Letting $A_{n,{\rm e}}$ denote the operator $A$ in this representation, 
we obtain the relation 
\begin{equation*}
A_{n,{\rm e}}= 
\left( \sum_{j=1}^n Q_j^2 \right)^{-1} \sum_{j=1}^n  Q_j P_j =
- i
 \left( \sum_{j=1}^n \frac{2 x_j}{r} 
 \frac{\partial}{\partial x_j} \right)   
\end{equation*}
with $r:= 2 \sum_{j=1}^n x_j^2$.
Now define the unitary map
$U_n: L^2(\real ^n) \to L^2(\real ^+) \otimes L^2(S^{n-1}) 
\cong L^2(\real ^+ \times S^{n-1})$ 
by 
$\Bigl(U_n(f)\Bigr) \bigl( r , ( e_1 , e_2, \ldots , e_n) \bigr) = 
r^{\frac{n-2}{4}} f( \sqrt{\frac{r}{2}}e_1,
\sqrt{\frac{r}{2}}e_2, \ldots ,\sqrt{\frac{r}{2}}e_n )$ , 
where $S^{n-1}$ denotes the $(n-1)$-dimensional spherical surface 
and $(e_1, e_2 , \ldots , e_n)$ is an element of $S^{n-1}$.
Then, the following relations hold;
\begin{eqnarray*}
\lo  U_n E_0 U_n^* = E_{0,\frac{n-2}{4}}\otimes I , \quad
U_n E_+ U_n^* = E_{+,\frac{n-2}{4}}\otimes I , \quad
U_n E_- U_n^* = E_{-,\frac{n-2}{4}}\otimes I ,\\
\lo U_n A_{n,{\rm e}} U_n^* =  
\left(P + i \left( \frac{n}{4} -\frac{1}{2} \right) Q^{-1}\right)
\otimes I= - i \frac{\partial}{\partial r} 
+ i \left( \frac{n}{4} -\frac{1}{2} \right)\frac{1}{r}
, \\
\lo U_n L^2_{\rm e}(\real^n)= L^2(\real^+) \otimes \psi_n,  \\
\lo U_n \odom(A_{n,{\rm e}}) \cap \left( C^1(\real^+) \otimes \psi_n \right) \\
=
\left\{ x^{\frac{n}{4} -\frac{1}{2}} f(x) 
\in  L^2(\real^+) \cap C^1(\real^+)
\left|
\begin{array}{l}
x^{\frac{n}{4} -\frac{1}{2}} f'(x) \in L^2(\real^+) \\
f(s) \,< \infty \hbox{ as } s \to 0 .
\end{array}
\right. \right\}
\otimes \psi_n ,
\end{eqnarray*}
where $\psi_n$ denotes the constant function on $S^{n-1}$ 
such that $\| \psi_n\|=1$.
The compound-system-type normal extension 
of $A_{n,{\rm e}}$ in the above 
relations is reduced to the discussion 
of $A_{\frac{n}{4} -\frac{1}{2}}$ which will be treated 
in section \ref{s51} and section \ref{s52}.

\section{Construction of compound-system-type normal extension of $A^*$}
\label{s5}

\subsection{The case where $\lambda =1$}\label{s51}
In this subsection, we will construct 
an compound-system-type normal extension of $A^*$ when $\lambda =1$. 
Let $\{ | \uparrow \rangle, | \downarrow \rangle \}$ be 
a CONS of $\complex^2$. From Lemma \ref{H10} and the fact that
$A^*$ is maximal symmetric, we obtain the following theorem.
\begin{thm}
Define the operator $T:= A \otimes | - \rangle \langle + | + 
A^* \otimes | + \rangle \langle - |$
on the domain $\odom(T):= \odom(A) \otimes | + \rangle \oplus \odom(A^*) \otimes | - \rangle$ with
$| \pm \rangle := \frac{1}{\sqrt{2}} (|  \uparrow\rangle \pm | \downarrow
 \rangle)$. 
The operator $T$ is a self-adjoint operator.
Moreover, the triple 
$(\complex^2, T  , | \uparrow \rangle )$ is a compound-system-type 
normal extension of $A^*$.
\end{thm}
Similarly, we can construct a compound-system-type normal extension of $a^*$
according to Lemma \ref{H10.1}.
The spectrum of the compound-system-type normal extension 
of $A^*$ for $\lambda =1$ appears only on the real axis.
That of the compound-system-type normal extension of $a^*$ 
appears only on the unit circle. 
\subsection{The cases where $\lambda \,>1$}\label{s52}
In the following, we will discuss the cases when $\lambda \,>1$.
Let $\{ | \uparrow \rangle, | \downarrow \rangle \}$ be CONS of $\complex^2$.
We obtain the following theorem, 
with $A_0$ ($A_k$ with $k=0$) discussed at the end of section \ref{s42}.
\begin{thm}
The pair of $E_+ \otimes \Id $ and $E_0 \otimes \Id + \Id \otimes E_0$
on ${\cal H}_{\lambda} \otimes {\cal H}_{\lambda -1}$
satisfies the commutation relation of the generators of 
Affine group.
This representation of Affine group is
written as follows;
there exist a Hilbert space ${\cal H}'$ and 
a unitary map $U$ from ${\cal H}_{\lambda} \otimes {\cal H}_{\lambda -1}$
to ${\cal H}' \otimes L^2(\real^+)$ such that
$U (E_+ \otimes \Id ) U^* = \Id \otimes E_+ ,
U ( E_0 \otimes \Id + \Id \otimes E_0 ) U^* = \Id \otimes E_0 $.
Then, the operator $U^* (\Id \otimes A_0) U \otimes | - \rangle \langle + | +
U^* (\Id \otimes A_0^*) U \otimes | + \rangle \langle - |$
with the domain $\odom \bigl( U^* (\Id \otimes A_0) U \bigr) \otimes | 
+ \rangle \oplus 
\odom\bigl( U^* (\Id \otimes A_0) U\bigr) \otimes | - \rangle$ is self-adjoint.

Moreover, the operator $T:=
U^* (\Id \otimes A_0) U \otimes | - \rangle \langle + | +
U^* (\Id \otimes A_0^*) U \otimes | + \rangle \langle - | 
- i E_+^{-1} \otimes E_+ \otimes \Id$ with the domain 
$\odom(T):= \left( \odom\bigl( U^* (\Id \otimes A_0) U \bigr) \otimes | 
+ \rangle \oplus 
\odom\bigl( U^* (\Id \otimes A_0^*) U\bigr) \otimes | - \rangle \right)
\cap \odom(  E_+^{-1} \otimes E_+) \otimes \complex^2$ is normal.
The triple $({\cal H}_\lambda':= {\cal H}_{\lambda-1} \otimes 
\complex^2, T , \psi :=  
| 0 \rangle_N  \otimes| \uparrow \rangle)$ 
is a compound-system-type normal extension of $A^*$.
\end{thm}
\begin{pf}
We need the discussion of Remark \ref{720.4} for the proof.
It is sufficient to prove them under the representations given in
section \ref{s42} because of Theorem \ref{720.3}.
Now define the unitary operator $U$ 
on $L^2( \real^+ ) \otimes L^2 (\real^+) $ by 
$\bigl(U(f)\bigr)(u,v) = \sqrt{v}f(v , uv)$. 
Then we have
$U (E_+ \otimes \Id ) U^* = \Id \otimes E_+ $,
$U (E_0 \otimes \Id + \Id \otimes E_0 ) U^* = \Id \otimes E_0 $
and $U (- i E_+^{-1} \otimes E_+ ) U^* = - E^+ \otimes \Id$.
Because the discussion at the end of section \ref{s42} shows that 
$A_0$ is closed and symmetric, 
it follows from the proof of Lemma \ref{H10}
that the operator 
$ A_0 \otimes | - \rangle \langle + | +
A_0^* \otimes | + \rangle \langle - | $ is self-adjoint
and its domain is $\odom(A^*) \otimes | - \rangle
\oplus \odom(A) \otimes | + \rangle$.
In general, for a self-adjoint operator $X$ on ${\cal K}_1$ and
a skew-adjoint operator $Y$ on ${\cal K}_2$,
we can show that the operator $X \otimes \Id + \Id \otimes Y$ with the domain 
$\odom(X) \otimes \odom(Y)= \odom(X) \otimes {\cal K}_2 \cap {\cal K}_1 \otimes 
\odom(Y) \subset {\cal K}_1 \otimes {\cal K}_2$ is normal.
Then, the operator $T':= \Id \otimes \left( A_0 \otimes | - \rangle
\langle +| \oplus A_0^* \otimes  | + \rangle \langle -| \right) 
- E^+ \otimes \Id \otimes \Id$ 
with the domain
$\odom(T') := \left( \odom( \Id \otimes A_0 ) \otimes | + \rangle \oplus 
\odom( \Id \otimes A_0^* ) \otimes | - \rangle \right)
\cap \odom(  E_+ ) \otimes L^2(\real^+) \otimes \complex^2$ is normal.
Thus, we have proved that the operator $T (= U^* T' U)$ is normal.
Now, we will prove that the triple $({\cal H}_\lambda',
T, \psi = | 0 \rangle_N^{k-\frac{1}{2}} \otimes | \uparrow \rangle )$ 
is a compound-system-type normal extension of $A_k^*$.
Since the set $\odom(A_k^*) \cap C^1(\real^+)$ is a core of the operator 
$A_k^*$,
it is sufficient to show that 
$(A_k^* \phi ) \otimes | 0 \rangle_N^{k-\frac{1}{2}} 
\otimes | \uparrow \rangle =
T \left( \phi \otimes | 0 \rangle_N^{k-\frac{1}{2}}
 \otimes | \uparrow \rangle \right)$
for any $\phi \in \odom(A_k^*) \cap C^1(\real^+)$.
\par From the definitions and (\ref{11.2}), some calculations result in 
\begin{eqnarray*}
\fl U \left( \left( \odom(A_k^*) \cap C^1( \real^+) \right) \otimes 
| 0 \rangle_{N}^{k-\frac{1}{2}} \right) \\
\lo= \{ f(v)u^{k-1/2}e^{-uv} \in L^2( \real^+ \times \real^+)|
x^{-k} f'(x) \in L^2(\real^+), 
f(s) \to 0 \hbox{ as } s \to 0\} . 
\end{eqnarray*}
We can show that a function $u \mapsto u^{k-1/2}e^{-uv}$ is contained by
$\odom(E_+) \subset L^2(\real^+)$ for any $v \in \real^+$.
If a function $f$ satisfies the condition $x^{-k} f'(x) \in L^2(\real^+), 
f(s) \to 0 \hbox{ as } s \to 0$,
then a function $v \mapsto f(v)u^{k-1/2}e^{-uv}$ is contained by
$\odom(A_0^*) \subset L^2(\real^+)$ for any $u \in \real^+$.

Then, the set 
$U \left( \left( \odom(A_k^*) \cap C^1( \real^+) \right) \otimes 
| 0 \rangle_{N}^{k-\frac{1}{2}}\right)$ is included in  
the set\par\noindent $\odom(\Id \otimes A_0^* ) \cap \odom( E_+ \otimes \Id) 
\cap \left( C^1(\real^+) \otimes C^1(\real^+) \right)$.
Hence, 
\begin{eqnarray*}
\fl U \left( \left( \odom(A_k^*) \cap C^1( \real^+) \right) \otimes 
| 0 \rangle_{N}^{k-\frac{1}{2}} \right) \otimes | \uparrow\rangle \\
\lo{\subset} \odom(\Id \otimes A_0^*\otimes \Id ) 
\cap \odom( E_+ \otimes \Id\otimes \Id) 
\cap \left( C^1(\real^+) \otimes C^1(\real^+)
\otimes | \uparrow\rangle\right) \\
\lo{\subset}  
\odom(\Id \otimes ( A_0^*\otimes | + \rangle \langle -|
+ A_0 \otimes | - \rangle \langle + | ) )
\cap \odom( E_+ \otimes \Id\otimes \Id)  \\
\cap \left( C^1(\real^+) \otimes C^1(\real^+) 
\otimes | \uparrow\rangle\right) \\
\lo{=} \odom( T' )
\cap \left( C^1(\real^+) \otimes C^1(\real^+) 
\otimes | \uparrow\rangle
\right).
\end{eqnarray*}
Thus, for the function $f(x)$ satisfying $\phi(x)= f(x) x^{-k}$, 
we obtain 
\begin{eqnarray*}
\fl T (\phi \otimes | 0 \rangle_{N}^{k-\frac{1}{2}} \otimes 
| \uparrow \rangle) \\
\eql - i \frac{\,d }{\,d v}
\left( f(v) u^{k-\frac{1}{2}} e^{-uv} \right) \otimes 
\left(| + \rangle \langle -| + | - \rangle \langle +|\right)
| \uparrow \rangle 
- i f(v) u^{k +\frac{1}{2}} e^{-uv} \otimes | \uparrow \rangle \\
\eql 
- i \frac{\,d f}{\,d v}(v) u^{k-\frac{1}{2}} e^{-uv} 
\otimes | \uparrow \rangle \\
\eql (A_k \phi ) \otimes | 0 \rangle_{N}^{k-\frac{1}{2}} \otimes | \uparrow 
\rangle.
\end{eqnarray*}
The theorem is now immediate.
\end{pf}
In the above discussions, it is sufficient only to choose 
${\cal H}_{\lambda -1}$ instead of ${\cal H}_{\lambda}'$ 
in order only to show that 
the operator $T$ formally satisfies $[ T,T^*]=0$ and 
formally satisfies (\ref{99.1}). 
However, the above definition of ${\cal H}_{\lambda}'$ is 
required in order that $T$ may be a normal operator 
defined in Definition \ref{D1}.

Since the spectrum of the compound-system-type normal extension 
of $A^*$ for $\lambda =1$ appears only in the upper half plane 
including the real axis,
the spectrum of the compound-system-type normal extension of $a^*$ 
appears only on the unit disk (including the unit circle) 
if the latter is related to the 
former by the adjoint of the Cayley transform.

\section{Conclusions}
We have discussed subnormal operators as a class 
of generalized observables.
A POVM of a subnormal operator defined in Definition \ref{14.1}
has little information about its implementation.
However, in order to
describe not only the probability distributions characterized 
by the POVMs but also a framework of their implementations, 
we have defined compound-system-type normal extensions in section \ref{14.2}.
(The heterodyne measurement known in quantum optics is interpreted 
as a special case of compound-system-type normal extensions.) 
In these contexts, we have constructed 
the compound-system-type normal extensions of
two subnormal operators $a^*$ and $A^*$ canonically introduced 
from an irreducible unitary representation of $\su(1,1)$,
when the minimum eigenvalue $\lambda$ of the 
generator $L_0$ is not less than one.  The squeezed states are 
regarded as the coherent states of the algebra $\su(1,1)$, 
and have been characterized as the eigenvectors of an operator 
defined in this mathematical framework.  The squeezed states in 
two-particle or multi-particle systems have been interpreted 
as the eigenvectors of the adjoints $a$ and $A$ of the subnormal 
operators $a^*$ and $A^*$. The 
coherent states of the affine group have been interpreted 
in the same framework, as well.  The squeezed states 
in one-particle system have been interpreted 
as the eigenvectors of the operator $a$ and $A$, 
though the operators $a^*$ and $A^*$ are not subnormal 
and their compound-system-type normal extensions 
do not exist in this case because $\lambda$ is less than one 
in this case. 

The information described by a compound-system-type 
normal extension isn't enough to completely specify the experimental 
implementation, where the measurement of the normal operator 
on the compound system is performed by the measurement on each system 
after some interactions were made between the basic system 
and the ancillary system. Therefore, the formulation including this 
specification is one of future problems.
As another possibility, 
since the affine group is closely related to Poincar\'{e} group,
our results about the affine group 
may be applicable to the relativistic quantum mechanics.

\section*{Acknowledgments}
The authors would like to express our thanks to Dr. S. Shirai,
Dr. T. Mine and Mr. M. Miyamoto for their advice 
and comments on functional analysis.
They are indebted to Professor M. Ozawa for suggesting the references
\cite{Sz,RieszN}.
They thank Professor H. Nagaoka 
for useful discussions.
They are also grateful to thank the third referee for 
useful comments.
They also thank Professor F. Hiai for suggesting the originality of
the bounded version of Lemma \ref{H6}.
\section*{Appendix}
The following lemma about spectra is well-known.
(See Hiai and Yanagi \cite{HY}.)
In Hiai and Yanagi \cite{HY},
it is proved in the case of bounded operators.
But, it can be easily extended to the case of unbounded operators.
\begin{lem}\label{HA1}
For a densely defined operator $A$ on ${\cal H}$,
Let $\sigma_p(A),\sigma_c(A)$ and $\sigma_r(A)$,
be the point spectrum, the continuous spectrum 
and the residual spectrum, respectively.
Then we have the following relations:
\begin{itemize}
\item $\lambda \in \sigma_r(A) \Rightarrow \lambda^* \in 
\sigma_p(A^*)$ 
\item $\lambda \in \sigma_p(A) \Rightarrow \lambda^* \in 
\sigma_r(A^*) \cup \sigma_p(A^*) $
\item $\lambda \in \sigma_c(A) \Rightarrow \lambda^* \in 
\sigma_c(A^*) $.
\end{itemize}
\end{lem}

\end{document}